\documentclass{raa}
\usepackage{graphicx,times}             %for PS/EPS graphics inclusion, new
\usepackage{natbib}
\usepackage{amssymb,amsmath}          %for PS/EPS graphics inclusion, new
\bibpunct{(}{)}{;}{a}{}{,}
\graphicspath{{}}
\usepackage[a4paper=true,pagebackref=true]{hyperref}
\hypersetup{colorlinks = true, linkcolor = green, anchorcolor = red, citecolor = blue, filecolor = red, pagecolor = red, urlcolor = red}

\begin{document}
	
   \title{A Cold and Diffuse Giant Molecular Filament in the Region of $l=41\dg$, $b=-1\dg$}
%   \subtitle{I. Place Your Subtitle Here}
   \volnopage{Vol.0 (20xx) No.0, 000--000}      %%preserved for Editor. DOn't remove!
   \setcounter{page}{1}          %%starting page, preserved for Editor. DOn't remove!
   \author{ Liang-Hao Lin
		\inst{1,2}
	\and Hong-Chi Wang 
		\inst{1,2}
	\and Yang Su
		\inst{1}
	\and Chong Li
		\inst{1,3}
	\and Ji Yang
		\inst{1,2}
   }
%	\institute{National Astronomical Observatories, Chinese Academy of Sciences,
%             Beijing 100012, China; {\it aiying@bao.ac.cn}\\
	\institute{
		Purple Mountain Observatory and Key Laboratory of Radio Astronomy, Chinese Academy of Sciences, 10 Yuanhua Road, Qixia District, Nanjing 210033, P.R. China; {\it lhlin@pmo.ac.cn}\\
        \and
        School of Astronomy and Space Science, University of Science and Technology of China, 96 Jinzhai Road, Hefei 230026, P.R. China
        \and
        University of Chinese Academy of Sciences, 19A Yuquan Road, Shijingshan District, Beijing 100049, P.R. China
		}
\vs\no
   {\small Submitted~~2019 Oct 08; Revised~~2020 March 31; Accepted~~2020~~April 08}

\abstract{
Data of $^{12}$CO/$^{13}$CO/C$^{18}$O $J=1\to0$ emission toward the Galactic plane region of $l=35\dg$ to $45\dg$ and $b= -5\dg$ to $+5\dg$ are available with the Milky Way Imaging Scroll Painting (MWISP) project.
Using the data, we found a giant molecular filament (GMF) around $l\approx38\sim42\dg$, $b\approx-3.5\sim0\dg$, $V_{LSR} \approx 27 \sim 40$ km~s$^{-1}$, named the GMF MWISP G041-01. 
At a distance of 1.7 kpc, the GMF is about 160 pc long.
With a median excitation temperature about 7.5 K and a median column density about $10^{21}$ cm$^{-2}$, this GMF is very cold and very diffuse compared to known GMFs.
Using the morphology in the data cube, the GMF is divided into four components among which three show filamentary structure.
Masses of the components are $ 10^3 \sim 10^4 M_\odot$, with a total mass for the whole filament being about $7\times10^4 M_\odot$ from the LTE method.
$^{13}$CO cores inside each component are searched.
Virial parameters are about 2.5 for these cores and have a power-law index of -0.34 against the mass.
The mass fraction of dense cores traced by $^{13}$CO to the diffuse clouds traced by $^{12}$CO are about 7\% for all components of the GMF.
We found signatures of possible large scale filament-filament collision in the GMF.
}
%
%   \authorrunning{L.-H. Lin et al. }            %author_head in even pages
   \titlerunning{Giant Molecular Filament MWISP G041-01}  % title_head in odd pages
   \maketitle
%% The author head (on even pages) and the title head (on odd pages) will be
%% automatically extracted from \author{} and \title{}. Whenever the title is too long,
%% you will be asked to supply a shorter one by inserting either \authorrunning{} or
%% \titlerunning{} before \maketitle. Anyway, you can specify your own heads.
%%
%% Note: In the following text body of your manuscript, please note several differences from
%%       other major journals:
%% (1) \subsection{Please Capitalize the First Letter of Each Notional Word in Subsection Title}
%% (2) Please Capitalize the First Letter of Each Notional Word in all tables' captions

%
%________________________________________________ sections below
%
\section{Introduction}           %% first-level sections will be auto-capitalized
\label{sect:intro}
% GMC
Giant molecular clouds (GMCs) are the sites where most of star formation take place and they are the largest gravitationally bound objects in the Milky Way \citep{2015ARA&A..53..583H}.
Results from \textit{Herschel} telescope have shown that filamentary structures in pc scale are ubiquitous in molecular clouds and they play an essential role in the processes of star formation \citep{2014prpl.conf...27A}.
% GMF
Discovered by \citet{2010ApJ...719L.185J} and later studied by \citet{2014ApJ...797...53G}, the Nessie cloud is the first giant molecular filament (GMF) found to have a length $\sim100$ pc.
Soon later, some unbiased surveys were carried out to search for GMFs based on surveys in continuum and spectral lines.
Those works were based on different data sets, with initial selection from continuum maps and velocity coherence checked later.
\citet{2014A&A...568A..73R} used the \textit{Spitzer}/GLIMPSE maps, while \citet{2015MNRAS.450.4043W} used the Hi-GAL maps, and both works checked the velocity coherence using the $^{13}$CO line data from the Galactic Ring Survey \citep[GRS,][]{2006ApJS..163..145J}.
\citet{2015ApJ...815...23Z} made the selection on the basis of the \textit{Spitzer}/MIPSGAL maps with additional velocity check using the following five surveys: HOPS, MALT90, BGPS follow-up, GRS and ThrUMMS.
\citet{2016ApJS..226....9W} searched for GMFs on the basis of the Bolocam Galactic Plane Survey (BGPS) sources and additional velocity information from follow-up HCO$^+(3\to2)$ and N$_2$H$^+(3\to2)$ observation.
Physical properties of GMFs and other similar structures are investigated by \citet{2018ApJ...864..153Z} and their star-forming contents are analysed by \citet{2019A&A...622A..52Z}, showing that GMFs are important both in tracing Galactic spiral structure and hosting star formation activities.
% CCC?

% Others Data
To date, most searches for GMFs started with continuum maps to identify absorption features in near/mid-infrared or emission features in far-infrared, and they are roughly limited to $|b|<1\dg$ due to the sky coverage of many Galactic survey programs \citep{2018ApJ...864..153Z}.
Previous searches for GMFs are based on dust filament candidates from continuum maps, and therefore they have higher column density thresholds than with the CO emission data.
The Milky Way Imaging Scroll Painting (MWISP) project has completed a $^{12}$CO/$^{13}$CO/C$^{18}$O $J=1\to0$ survey toward the Galactic plane region $l=35\dg$ to $45\dg$ and $b= -5\dg$ to $+5\dg$ \citep{2016ApJ...828...59S, 2019ApJS..240....9S}, providing the opportunity to search for GMFs at relatively high Galactic latitudes directly from the molecular emission for the first time.
Using the data, we found a GMF which extends from $( l = 42.5\dg, b=-0.5\dg ) $ to $ (l = 38.5\dg, b=-2.5\dg)$ and has a local stand of rest (LSR) velocity between 27 and 40 km~s$^{-1}$, which we designate as GMF MWISP G041-01. 
This GMF was also named as GMF G40.82-1.41 in Section 3.3.1 of \citet{2018ApJ...863..103S}.

% Old data
Though several previous surveys have already covered this GMF in different wavelengths, it has not be analysed in detail.
For example, the survey by \citet{2001ApJ...547..792D} covered this GMF in  $^{12}$CO $J=1\to0$ line but it can not resolve the internal structure due to the low spatial resolution of the CfA 1.2 m telescope used for the survey.
In the \textit{Planck} derived CO map and the far-infrared thermal dust emission map, this GMF is barely detectable \citep{2016A&A...594A..10P}.
This GMF is visible in the near-infrared absorption map of Figure 5 in \citet{2018MNRAS.478..651G}.
We derived a dust distance modulus of $10.7\pm0.2$ (i.e. $1.4\pm0.2$ kpc) toward this GMF using the result cube of \citet{2018MNRAS.478..651G}.
% Distance
The kinematic distance is $ 2.0\pm0.4 $ kpc based on the A5 model of \citet{2014ApJ...783..130R}.
We use the average, 1.7 kpc, of the dust distance and the kinematic distance for this GMF in this paper.
% No Star Forming
Neither \hii ~regions nor supernova remnants are found to be associated with this GMF in the radio contiunum map \citep{1998AJ....115.1693C} and catalogs \citep{2014ApJS..212....1A, 2019JApA...40...36G}.
The distribution of young stellar objects(YSOs) derived from \textit{AllWISE} does not show obvious enhancement within this GMF \citep{2016MNRAS.458.3479M}.
These facts indicate that this GMF does not host any apparent star formation activity.

The paper is arranged as follows.
In section \ref{sect:obs}, we describe our observations and the data used in this study.
In section \ref{sect:result:GMF+fila}, the entire GMF and its filamentary sub-structures are analysed.
In section \ref{sect:result:para}, physical parameters are derived for the GMF and each sub-structures.
In section \ref{sect:result:cores}, we present the results of molecular core searching inside the GMF.
In section \ref{sect:discussion}, we discuss the relation between the GMF and spiral arms, and a possible filament-filament collision (FFC) scenario of two filamentary components in the GMF.
We make a summary in section \ref{sect:Summary}. 

\section{Observations}
\label{sect:obs}
The MWISP project is a $^{12}$CO/$^{13}$CO/C$^{18}$O $J=1\to0$ survey of the northern Galactic plane which began in 2011 and now is still underway.
The project details and some initial results can be found in \citet{2019ApJS..240....9S}. 
For convenience, we list some main settings below.
It uses the 13.7 m millimetre-wave telescope located in Delingha, China.
The telescope is equipped with a $3\times3$ multibeam Superconducting Spectroscopic Array Receiver (SSAR) \citep{2012ITTST...2..593S} as the front end and a Fast Fourier Transform Spectrometer (FFTS) with a total bandwidth of 1 GHz allocated to 16384 channels as the back end.
Data used in this study cover the sky region of $l = 38 \sim 44\dg$, $b = -3.5 \sim +1\dg$, which was mapped from the autumn of 2011 to the spring of 2015 \citep{2016ApJ...828...59S}.

System temperature was about 250 K for the $^{12}$CO $J=1\to0$ line and 140 K for the $^{13}$CO and C$^{18}$O $ J=1\to0 $ lines while half-power beam width (HPBW) of the telescope was about 48$\arcsec$ for the $^{12}$CO line and 50$\arcsec$ for the $^{13}$CO and C$^{18}$O lines during the observing period.\footnote{\url{http://www.radioast.nsdc.cn/zhuangtaibaogao.php}}
The main beam efficiency was about 0.5 and the received signal has been converted to main beam brightness temperature $T_{MB}$ by the standard pipeline.
The velocity width per channel is 0.159 km s$^{-1}$ for the $^{12}$CO line and 0.166 km s$^{-1}$ for the $^{13}$CO and C$^{18}$O lines.
For easy comparison, data cubes for these three lines are re-binned to the same velocity channel of 0.2 km s$^{-1}$. 
The median $1\sigma$ noise is about 0.5 K for the $^{12}$CO line and 0.3 K for the $^{13}$CO and C$^{18}$O lines.

The data reduction were conducted using the GILDAS-CLASS\footnote{\url{http://www.iram.fr/IRAMFR/GILDAS/}} package.

\section{Results}
\label{sect:result}
\subsection{The Entire GMF and its Filamentary Structures}
\label{sect:result:GMF+fila}
% avspec + CO RGB
%-------	avspec
\begin{figure}
	\centering
	\includegraphics[width=\linewidth]{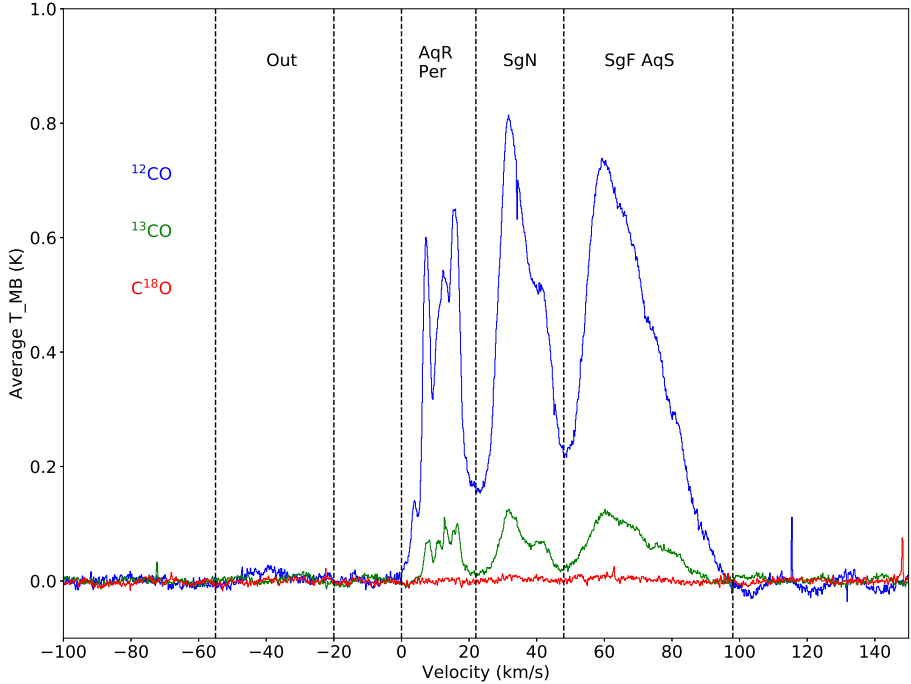}
	\caption{Average spectra of $^{12}$CO/$^{13}$CO/C$^{18}$O emission in the analysed region. Following \citet{2016ApJ...823...77R}, spiral arm segments and other features from near to far are labelled as: Aquila Rift (AqR), Sagittarius near portion (SgN), Aquila Spur (AqS), Sagittarius far portion (SgF), Perseus arm (Per) and Outer arm (Out).} 
	\label{Fig:avspec}
\end{figure}
%-----------------
The average spectra of $^{12}$CO/$^{13}$CO/C$^{18}$O emission in the analysed region are shown in Figure \ref{Fig:avspec}.
The GMF exhibits velocities in the range marked as the near portion of the Sagittarius arm.

%-------	CO_RGB
\begin{figure}
	\centering
	\includegraphics[width=\linewidth]{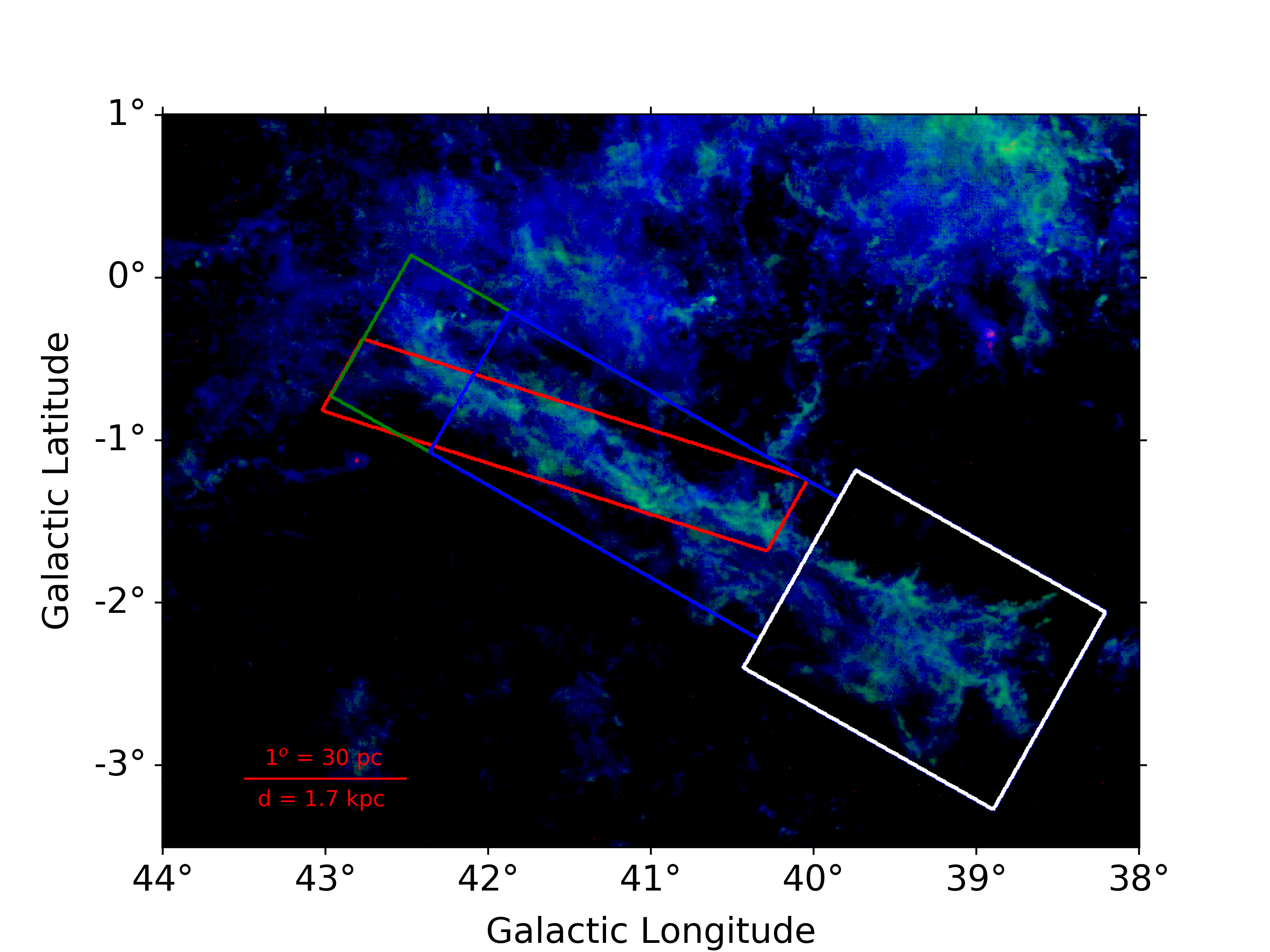}
	\caption{Integrated intensity map of the $^{12}$CO/$^{13}$CO/C$^{18}$O lines in the velocity range between 25 and 40 km s$^{-1}$ with $^{12}$CO in blue, $^{13}$CO in green and C$^{18}$O in red. C$^{18}$O line emission is weak and is concentrated in a few pixels. Boxes show boundaries of each component of the GMF, with F1, F2, F3 and SWP being outlined in blue, green, red and white, respectively.} 
	\label{Fig:CO_RGB}
\end{figure}
%-----------------
The integrated intensity map of the $^{12}$CO/$^{13}$CO/C$^{18}$O lines in the velocity range between 25 km s$^{-1}$ and 40 km s$^{-1}$ is shown in Figure \ref{Fig:CO_RGB}.
The GMF can be clearly distinguished from the surrounding clouds.
It was about 5$\dg$ in length and about 0.6$\dg$ in width, and has an angle about 30$\dg$ with respect to the $b=0\dg$ plane.
In our assumed distance of 1.7 kpc, the length is about 160 pc, and the largest separation from the Galactic plane is about 80 pc.
Both the large length and the relatively high latitude are unusual compared to other GMFs listed in \citet{2018ApJ...864..153Z}.
As the C$^{18}$O flux is barely detected, future discussion is limited to the $^{12}$CO and $^{13}$CO data.

% Channel map
%-------	U_channelmap
\begin{figure}
	\centering
	\includegraphics[width=\linewidth]{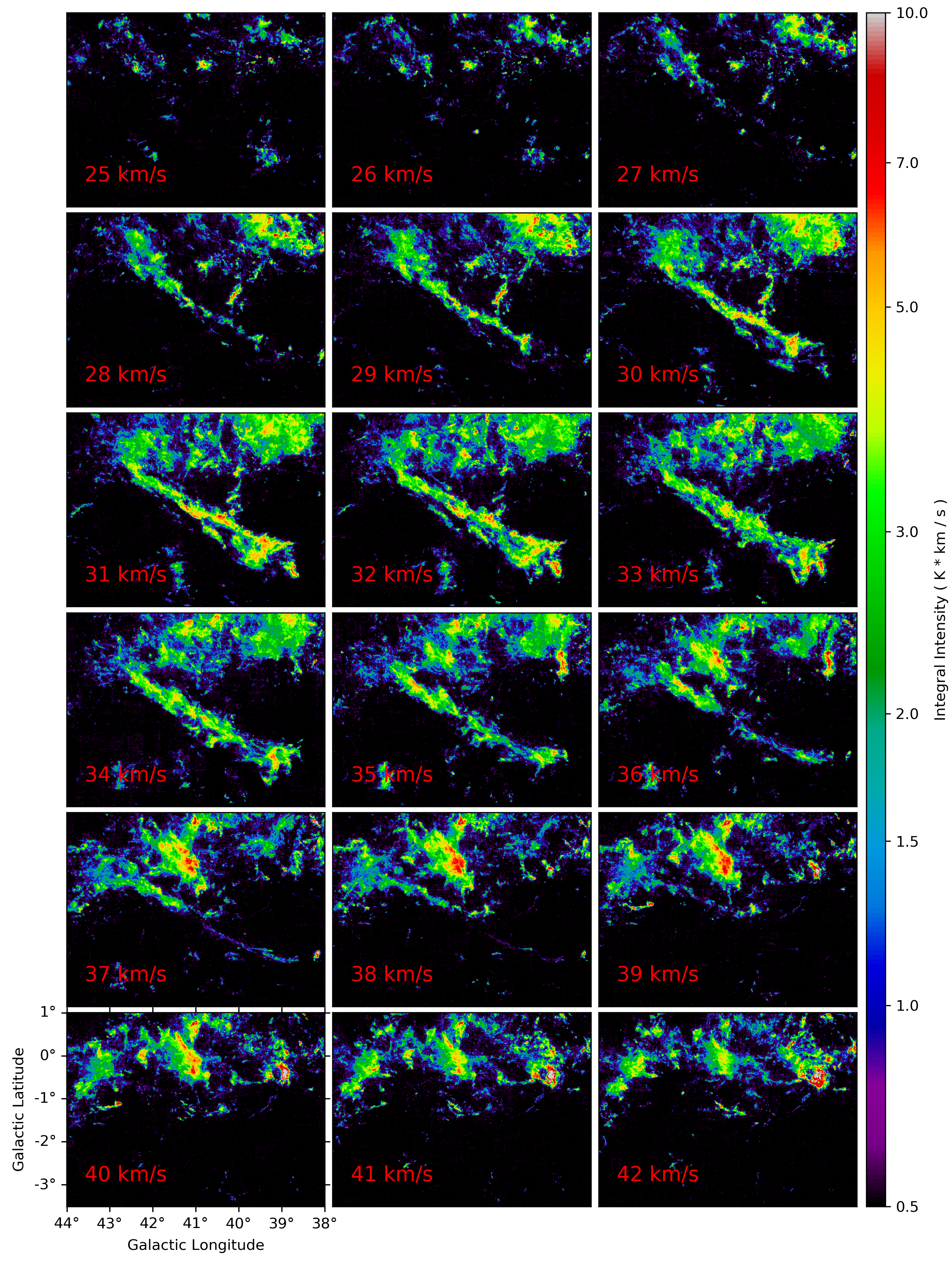}
	\caption{Channel map of $^{12}$CO line emission with the central velocity shown at the bottom-left corner of each panel.} 
	\label{Fig:U_channelmap}
\end{figure}
%-----------------
%-------	L_channelmap
\begin{figure}
	\centering
	\includegraphics[width=\linewidth]{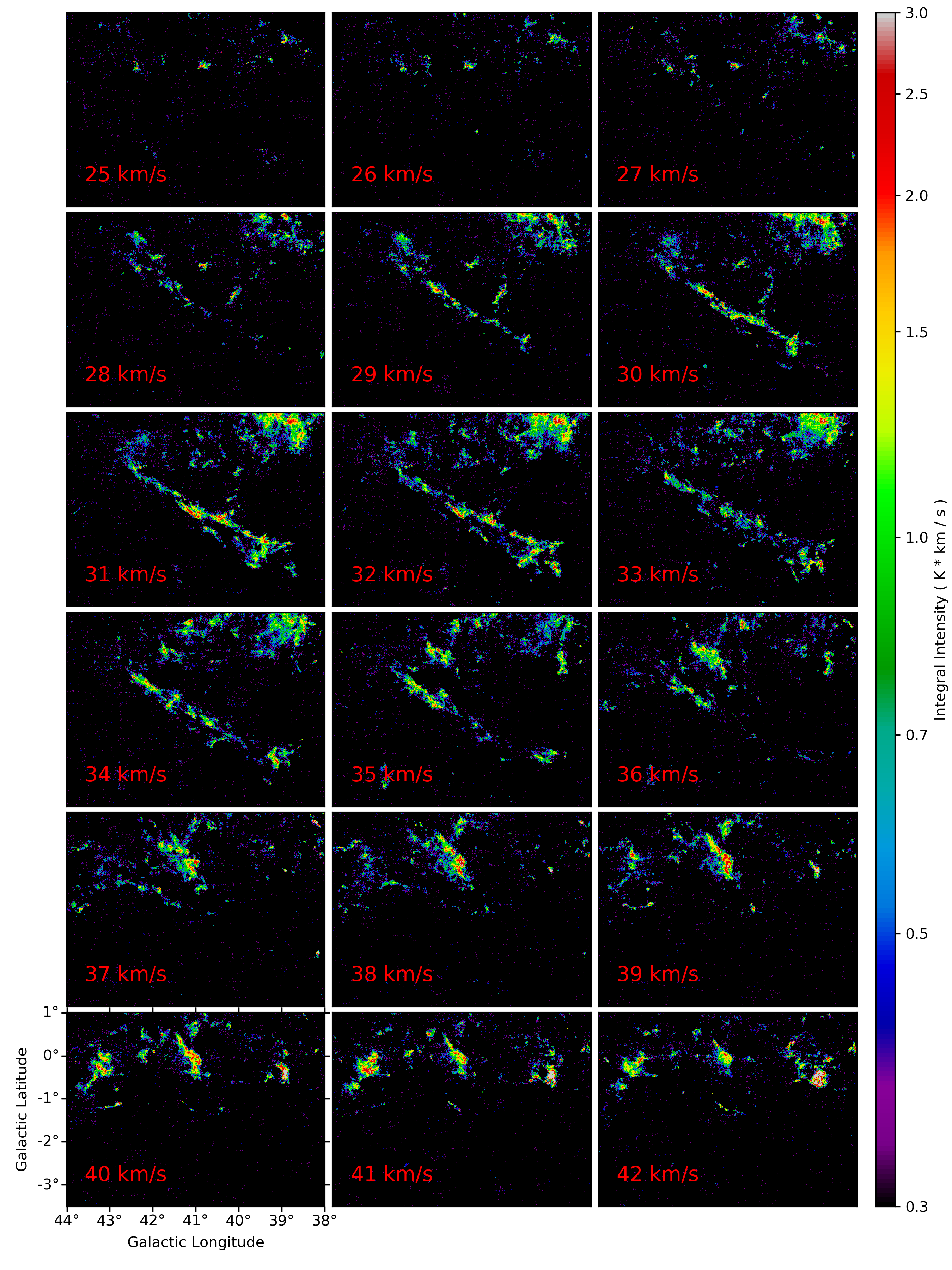}
	\caption{Same as Figure \ref{Fig:U_channelmap} but for the $^{13}$CO line emission.} 
	\label{Fig:L_channelmap}
\end{figure}
%-----------------
The channel maps, i.e. the integrated intensity map in 1 km s$^{-1}$ bins, of $^{12}$CO and $^{13}$CO lines are shown in Figures \ref{Fig:U_channelmap} and \ref{Fig:L_channelmap}, respectively.
The GMF lies from northeast to southwest in this region.
Clouds to the east of the GMF, around $(43.5\dg, -0.5\dg)$, have much higher LSR velocity and are farther than the GMF.
Clouds to the northwest, around $(39\dg, +1\dg)$, are the southern part of a much larger cloud that has a trigonometric distance about 1.9 kpc \citep{2014ApJ...783..130R}.
Cloud at around $(40\dg, 0\dg)$ is GMF 41 in \citet{2018ApJ...864..153Z}.
They placed it at 2.5 kpc.
However, the dust absorption data from \citet{2018MNRAS.478..651G} have provide a similar distance for GMF 41 as the GMF in this work.

%-------	L_m1
\begin{figure}
	\centering
	\includegraphics[width=\linewidth]{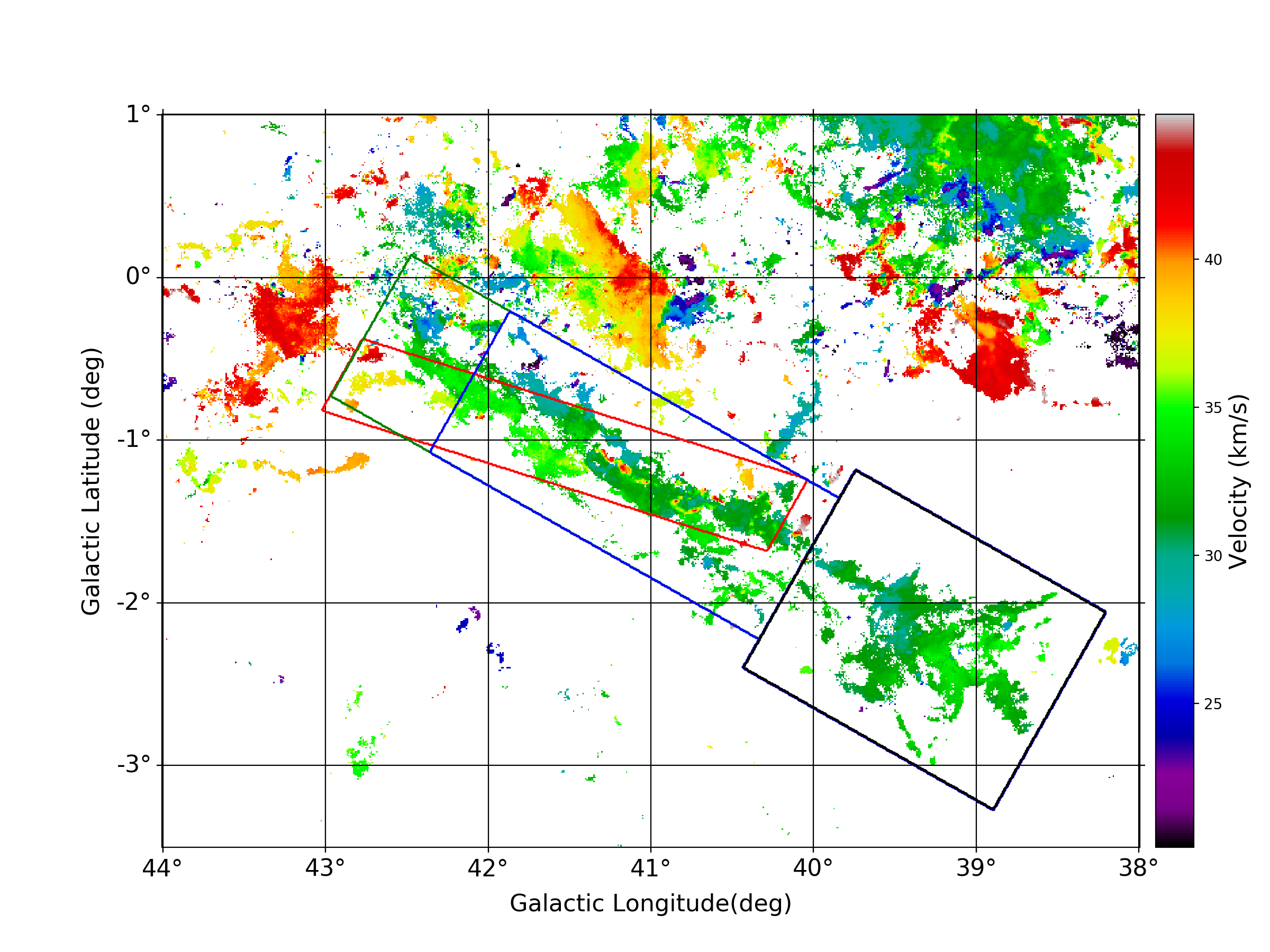}
	\caption{The intensity-weighted mean velocity map of $^{13}$CO line emission. We calculate the mean velocity in velocity range between 20 and 45 km s$^{-1}$ so that clouds with velocities higher than that of the GMF can be easily extinguished. The GMF has a different velocity from its neighbours. Boxes show boundaries of each component. F1, F2, F3 and SWP are in blue, green, red and black, respectively.}
	\label{Fig:L_m1}
\end{figure}
%-----------------
% L_m1 map
The moment-1 map, i.e. the flux-weighted mean velocity map, of the $^{13}$CO line emission for the analysed region is shown in Figure \ref{Fig:L_m1}.
The northeast part of the GMF has three velocity components, with velocities of 29 km s$^{-1}$, 34 km s$^{-1}$ and 38 km s$^{-1}$.
There is a significant velocity gradient in the south-west part (SWP) of the GMF along its main axis.

%-------	L_vpbelt
\begin{figure}
	\centering
	\includegraphics[width=\linewidth]{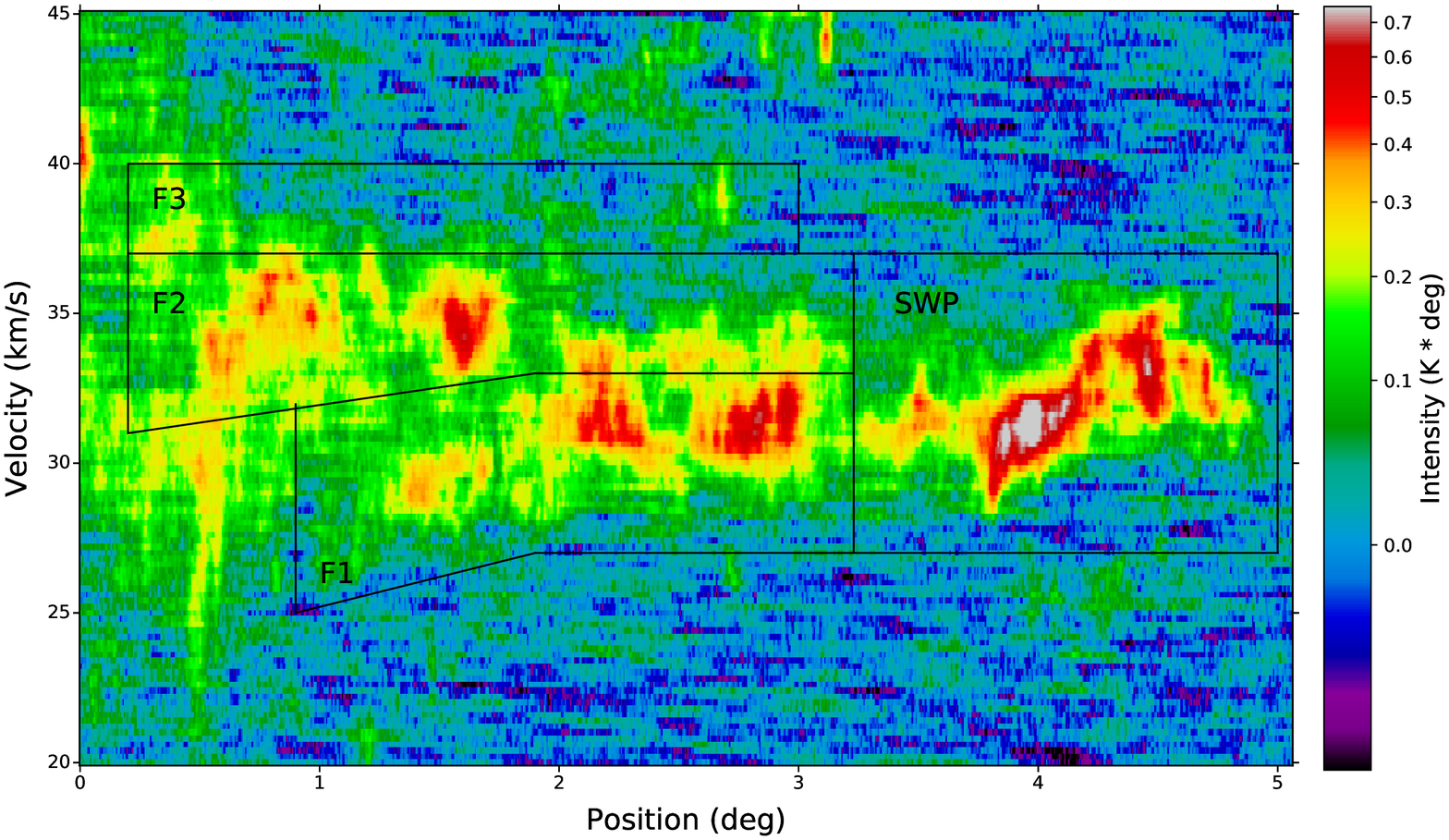}
	\caption{$^{13}$CO velocity-position diagram of the GMF. Position is from $(l,b)=(42.9\dg, -0.2\dg)$ to $(38.5\dg, -2.7\dg)$. Integral width is 120 pixel ($1\dg$). Boundary of each component of the GMF is shown with black line. F1 to F3 means filaments 1 to 3, and SWP means the southwest part of the GMF.}
	\label{Fig:L_vpbelt}
\end{figure}
%-----------------
% L_vpbelt map
To show the velocity structures of the GMF more clearly, a velocity-position map of $^{13}$CO line emission along the GMF is shown as Figure \ref{Fig:L_vpbelt}.
The position is from $(l,b)=(42.9\dg, -0.2\dg)$ to $(38.5\dg, -2.7\dg)$ and the integral width is 1 degree (about 30 pc).
Two main velocity components at 29 km s$^{-1}$ and 34 km s$^{-1}$ can be seen at the position offset around 1.5$\dg$.
The velocity gradient of the SWP is clearly shown in Figure \ref{Fig:L_vpbelt} and has a value of about 4 km s$^{-1}$ deg$^{-1}$, or 0.14 km s$^{-1}$ pc$^{-1}$.

% parts
Since these components are distinguishable both in space and velocity, we divide the GMF into three filamentary parts (F1 to F3) and one cloudy part (the SWP), and show their boundaries in Figure \ref{Fig:L_vpbelt}.

%-------	Fila3_U_m0
\begin{figure}
	\centering
	\includegraphics[width=\linewidth]{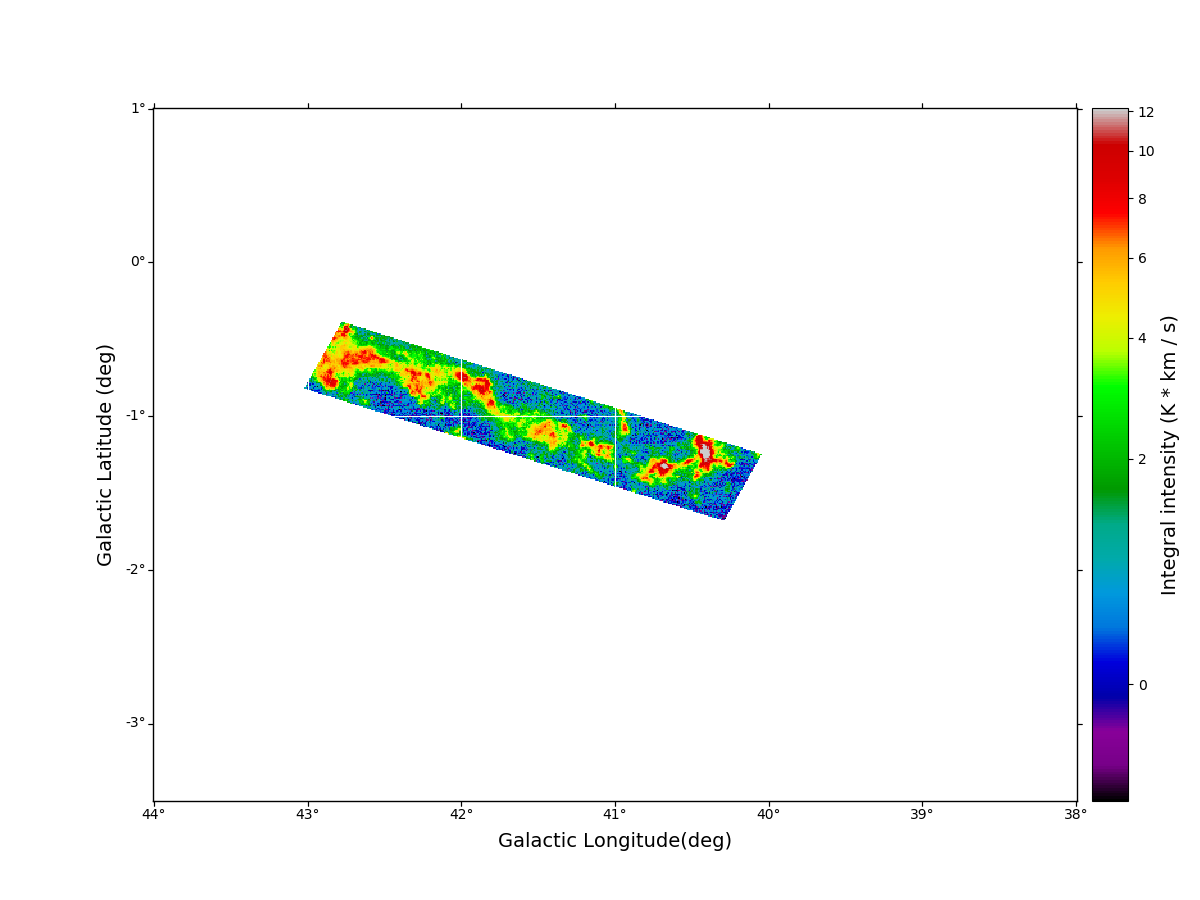}
	\caption{The $^{12}$CO integral intensity map of F3. Through it is not clear in the $^{13}$CO v-p map (Figure\ref{Fig:L_vpbelt}), the filament structure is clearly in this map with a "$\sim$" shape.}
	\label{Fig:F3_U_m0}
\end{figure}
%-----------------
As F3 is relatively weak in $^{13}$CO emission, its $^{12}$CO integral intensity is shown in Figure \ref{Fig:F3_U_m0}.
In this map, F3 shows a "$\sim$" shape and is fragmented into several clumps, with more clumps being located at its two ends.

\subsection{Physical Parameters of the Entire GMF and Each Components}
\label{sect:result:para}
%-------	T_ex & tau & N_H2
\begin{figure}
	\begin{subfigure}{0.33\linewidth}
		\centering
		\includegraphics[width=\linewidth]{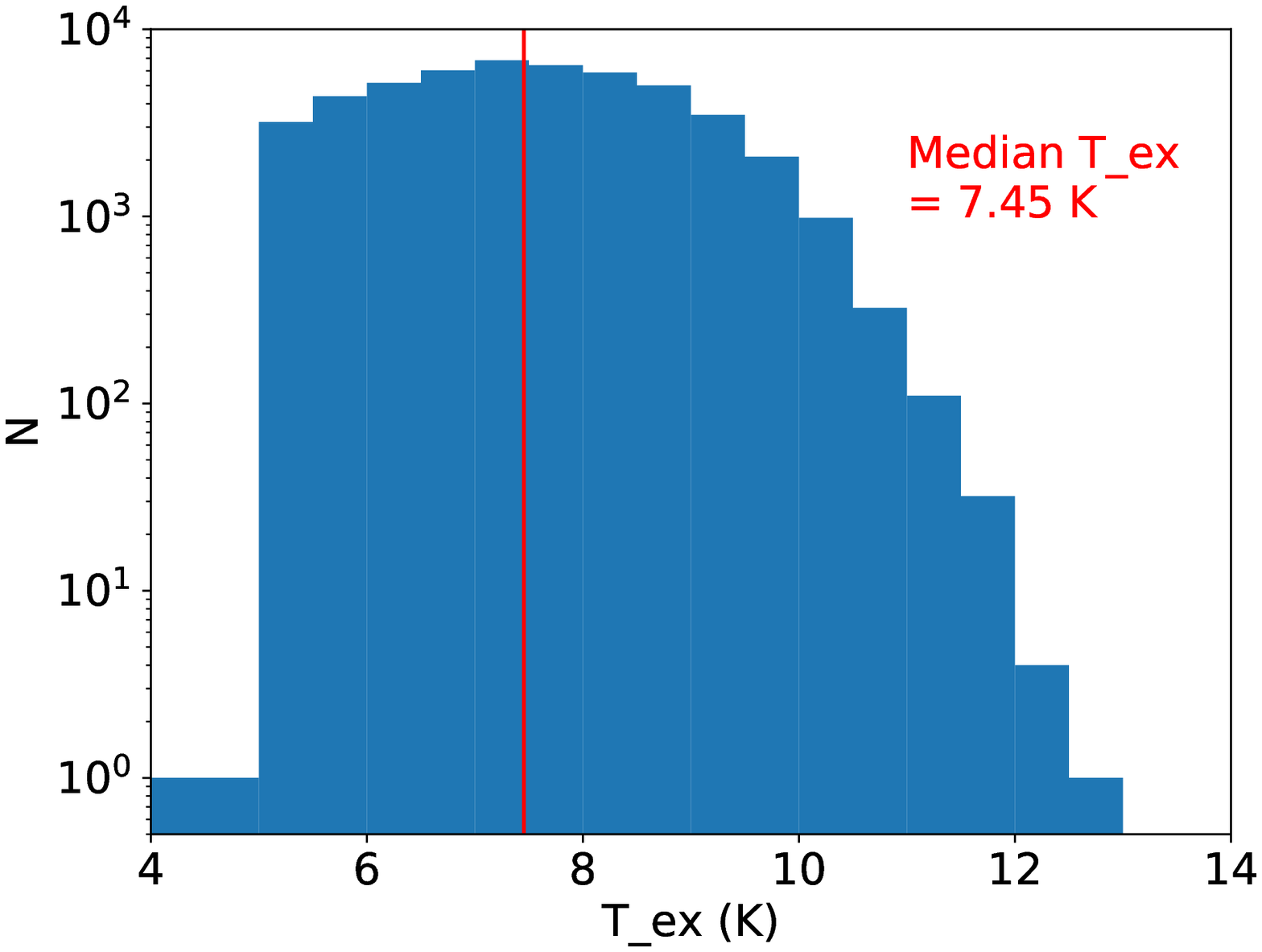}
		\caption{Excitation temperature $ T_{ex} $}
		\label{fig:Fila_Tex}
	\end{subfigure}
	\begin{subfigure}{0.33\linewidth}
		\centering
		\includegraphics[width=\linewidth]{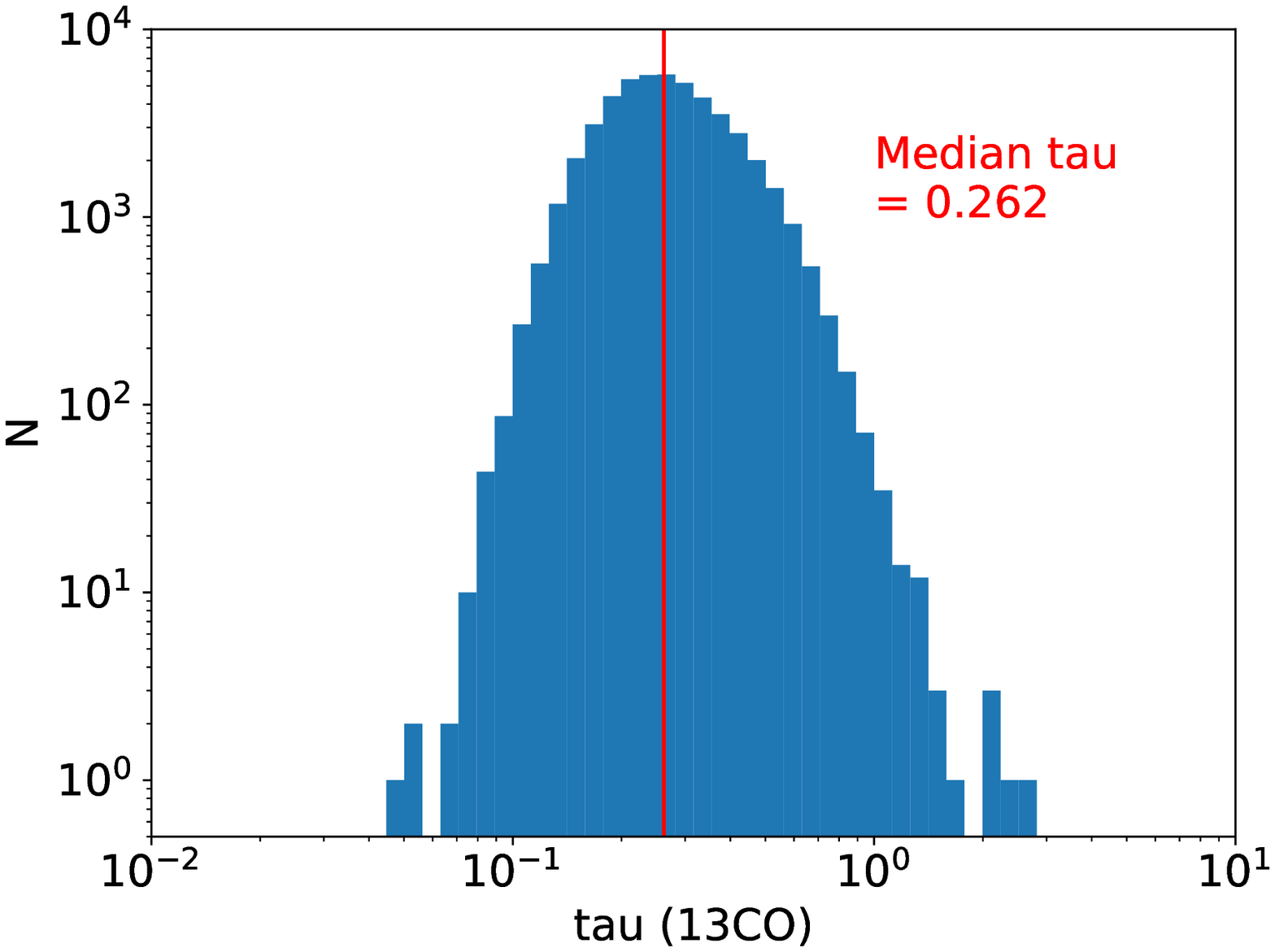}
		\caption{$^{13}$CO optical depth $\tau_{^{13}CO}$}
		\label{fig:Fila_tau}
	\end{subfigure}
	\begin{subfigure}{0.33\linewidth}
		\centering
		\includegraphics[width=\linewidth]{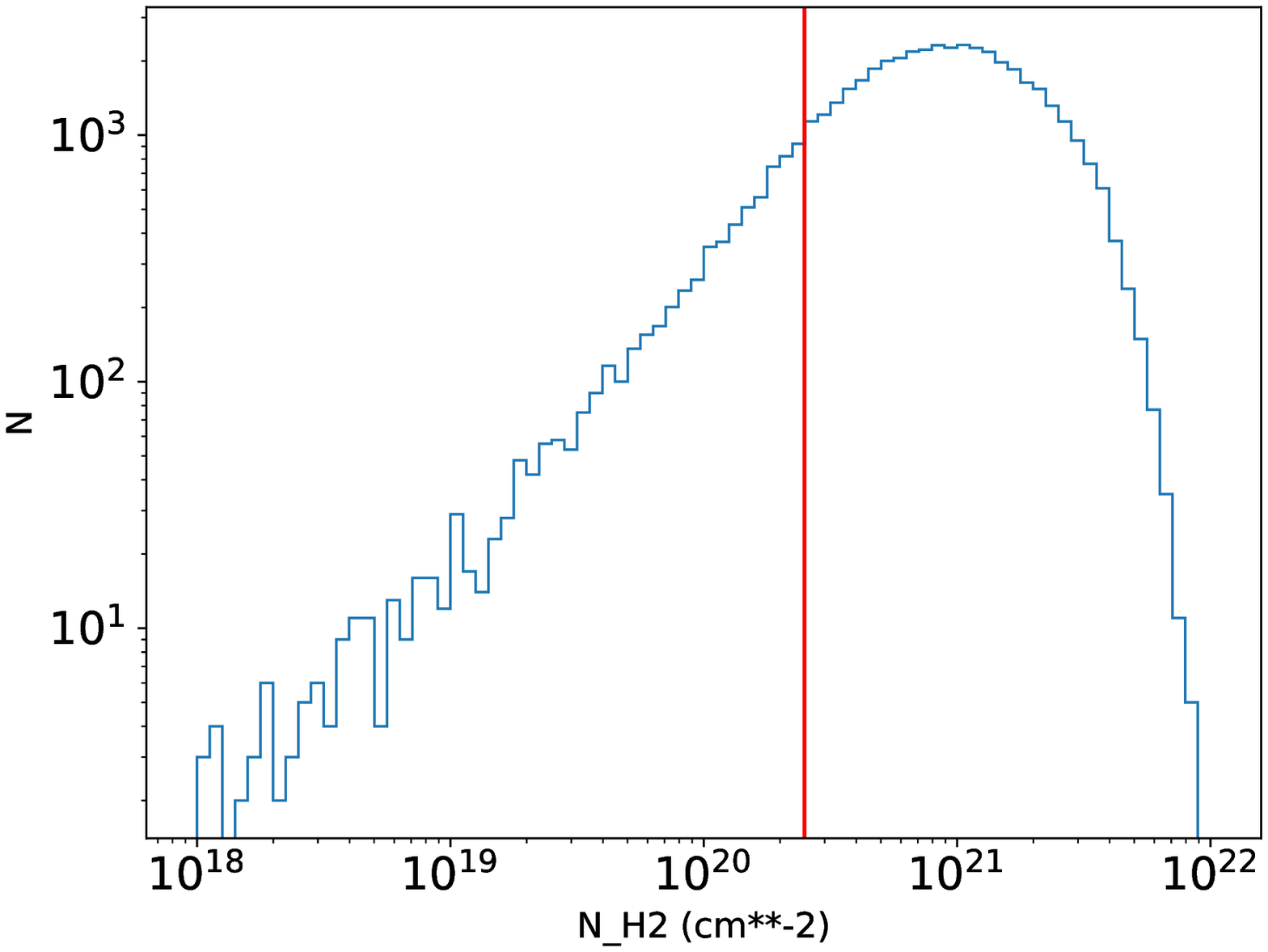}
		\caption{H$_2$ column density $N_{H_2}$}
		\label{fig:Fila_N}
	\end{subfigure}
	\caption{Distribution of (a) excitation temperature, (b) $^{13}$CO optical depth $\tau_{^{13}CO}$, and (c) H$_2$ column density $N_{H_2}$ for pixels in the GMF. Red line represent median value in (a) and (b), but detection limit in (c). The limit is calculated as $3\sigma$ $^{13}$CO emission within 3 contiguous channels with excitation temperature and optical depth taken to be the median values.}
	\label{Fig:Tex_tau_N(H2)}
\end{figure}
%------------------
The  molecular hydrogen gas column density $N_{H_2}$ can be derived by the X-factor method as
\begin{equation}
N_{\mathrm{H_2}} ( X_\mathrm{CO} ) = X_\mathrm{^{12}CO} {\int T_{MB}(^{12}\mathrm{CO}) dv}
\label{eq:N_H2(X_CO)}
\end{equation}
where $ X_{ ^{12}\mathrm{CO} } $ is the $^{12}$CO-to-H$_2$ conversion factor and we adopt the value of $2 \times 10^{20} $ molecules cm$^{-2}$ (K km s$^{-1}$)$^{-1}$ from \citet{2013ARA&A..51..207B}. $T_{MB}(^{12}\mathrm{CO})$ is the main beam brightness temperature of $^{12}$CO emission.

Another way to estimate $N_\mathrm{H_2}$ is the standard local thermodynamic equilibrium (LTE) radiation transfer method.
The $^{12}$CO line is assumed to be optically thick and the excitation temperature $T_{ex}$ is derived from the peak intensity of $^{12}$CO line $T_\mathrm{^{12}CO peak}$ as
\begin{equation}
J_{\nu_{12}}(T_{ex}) = k_B~T_\mathrm{^{12}CO peak} / h\nu_{12} + J_{\nu_{12}}(T_{bg})
\label{eq:T_ex}
\end{equation}
where $J_\nu(T) = \dfrac{1}{e^{h\nu/k_B T} - 1} $, $\nu_{12}=115271202$ kHz is the frequency of $^{12}$CO $J=1\to0$ line, $k_B$ is the Boltzmann constant, $T_{^{12}CO peak}$ is the peak intensity of the pixel and $T_{bg}=2.726$ K is the temperature of the cosmic microwave background.
Here we assumed the beam filling factor is 1 as the $^{12}$CO emmission is relatively extended.
This results in a derived excitation temperature of
\begin{equation}
T_{ex} = \dfrac{5.532~\rm K}{ln(\dfrac{5.532}{T_\mathrm{^{12}CO peak}/\rm K+0.837} + 1)}.
\label{eq:T_ex_value}
\end{equation}

Under the assumption of LTE, excitation temperatures for $^{12}$CO and $^{13}$CO are the same.
Then the optical depth of $^{13}$CO is 
\begin{equation}
\tau_\mathrm{^{13}CO} = -ln \left( 1 - \dfrac{k_B~T_{MB}(\mathrm{^{13}CO})}{h\nu_{13} } \dfrac{1}{J_{\nu_{13}}(T_{ex}) - J_{\nu_{13}}(T_{bg}) } \right)
\label{eq:tau}
\end{equation}
or equivalently 
\begin{equation}
\tau_\mathrm{^{13}CO} = -ln \left( 1 - \dfrac{T_{MB}(\mathrm{^{13}CO}) / 5.289~\rm K}{J_{\nu_{13}}(T_{ex}) - 0.168) } \right)
\label{eq:tau_value}
\end{equation}

Following \citet{2010ApJ...721..686P}, the column density of $^{13}$CO is
%\begin{equation}
%N_{^{13}CO} = \dfrac{8\pi k_B \nu_{13}^2}{hc^3A_{ij}} \dfrac{ 2k_B T_{ex} / h\nu_{13} } { 3 e^{-h\nu_{13}/k_B T_{ex}} } \dfrac{\tau_{^{13}CO}}{1-e^{-\tau_{^{13}CO}}} \dfrac{1}{1-e^{-hv_{13}/k_B T_{ex}}} \int T_{MB}(^{13}CO) dv
%\label{eq:N_13CO}
%\end{equation}
\begin{equation}
N_{^{13}\mathrm{CO}} = \dfrac{2.42 \times 10^{14} ~\rm cm^{-2}~K^{-1}~km^{-1}~s } { 1-e^{-5.289\rm K /\it T_{ex}} } \dfrac{\tau_{^{13}\mathrm{CO}}}{ 1- e^{-\tau_{^{13}\mathrm{CO}}} } \int T_{MB}(^{13}\mathrm{CO})dv
\label{eq:N_13CO_value}
\end{equation}
where $T_{MB}(\mathrm{^{13}CO})$ is the main beam brightness temperature of $^{13}$CO emission.
The number ratio for conversion used in this paper is $ f(\mathrm{H_2/^{12}CO}) = 1.1 \times 10^4 $ \citep{1982ApJ...262..590F} and $ f(\mathrm{^{12}C / ^{13}C}) = 77 $ \citep{1994ARA&A..32..191W}.
The H$_2$ column density is thus
\begin{equation}
N_{H_2}(LTE) = f(\mathrm{H_2/^{12}CO}) \times f(\mathrm{^{12}C / ^{13}C}) \times N_{^{13}\mathrm{CO}} \simeq 8.5 \times 10^5 \times N_{^{13}\mathrm{CO}}.
\label{eq:N_H2(LTE)}
\end{equation}

% T_ex & tau & N_H2
The distribution of excitation temperature $ T_{ex} $, $^{13}$CO optical depth $\tau_{^{13}CO}$, and H$_2$ column density $N_{H_2}$ for the whole GMF are shown in Figure \ref{Fig:Tex_tau_N(H2)}(a), (b) and (c), respectively.
The median excitation temperature is about 7.5 K.
The $^{13}$CO optical depth has a median value of 0.26.
The column density of this GMF is at the order of $10^{21}$ cm$^{-2}$, and the distribution does not follow a power-law relation at the high-density end.
Compared with the GMFs analysed by \citet{2018ApJ...864..153Z}, this GMF is much colder (7.5 K compared to 20.8 K) and about 5 times lower in column density.
The peak $T_{ex}$ of our GMF is smaller than 13 K, close to the lowest value of other GMFs.
We note that, both the excitation temperature and column density of our GMF are derived from CO spectroscopic data while these parameters for the known low-latitude GMFs are usually derived from SED fitting of dust continuum or ammonia spectra.
Due to the effect of the beam filling fact, our results for the excitation temperature of our GMF are underestimated while the column densities are overestimated because of using lower excitation temperature.
However, \citet{2019A&A...632A.115G} have pointed out that the kinematic temperature derived from RADEX non-LTE analysis for the L1188 filament (distance about 0.8 kpc) is about 13--23 K, not much higher than the excitation temperature (about 10--20 K) derived from the optical thick $^{12}$CO $J=1\to0$ line in their previous work \citep{2017ApJ...835L..14G}.
Keeping those caveats in mind, it is still reasonable to believe that this GMF is very cold and very diffuse compared to other GMFs in the literature.

\subsection{Cores in the GMF}
\label{sect:result:cores}
% Parameters for Cores

%-------	Para_core: v_rms, R, Tex
\begin{figure}
	\begin{subfigure}{0.45\linewidth}
		\centering
		\includegraphics[width=\linewidth]{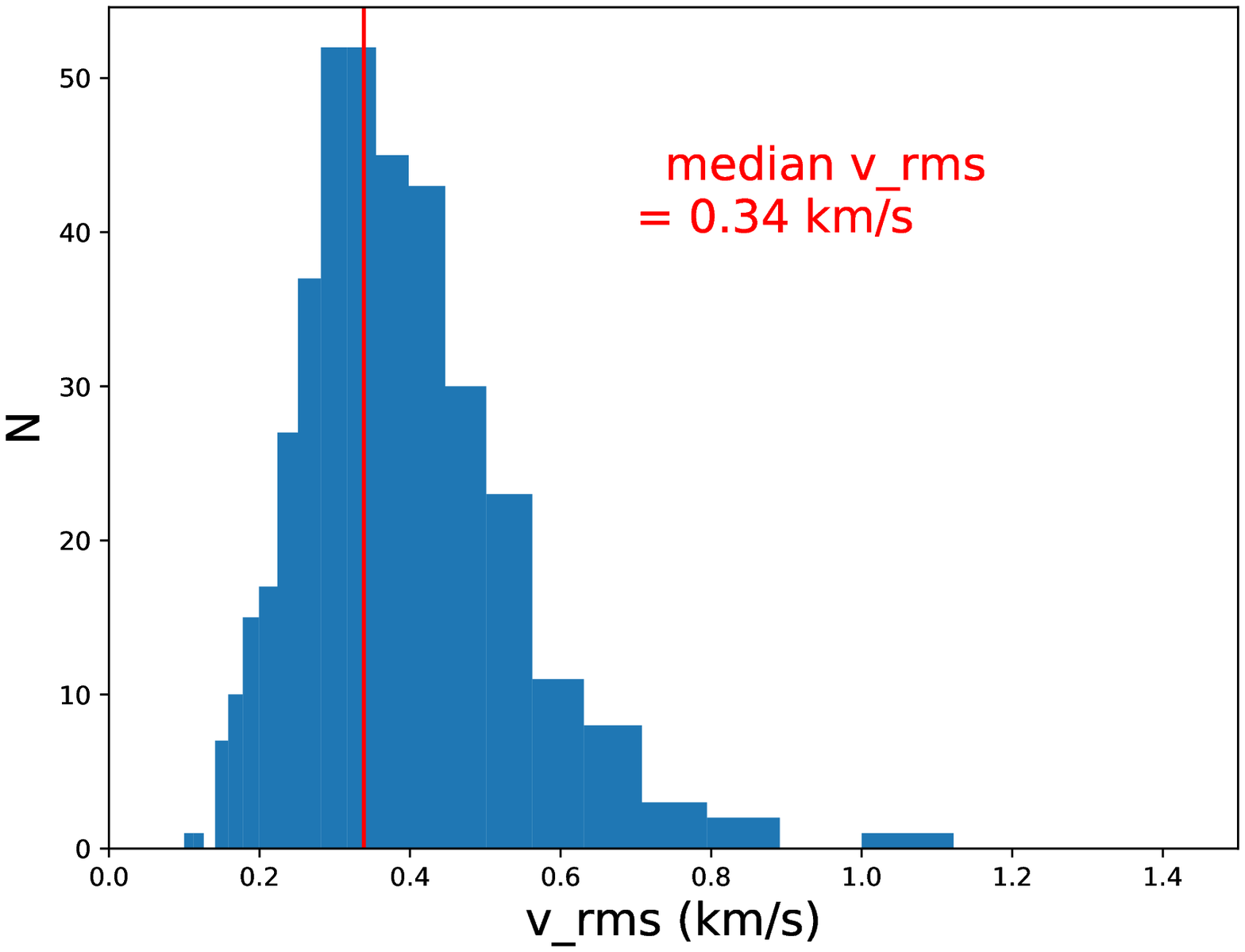}
		\caption{Velocity dispersion $v_{rms}$}
		\label{fig:core_v_rms}
	\end{subfigure}
	\begin{subfigure}{0.45\linewidth}
		\centering
		\includegraphics[width=\linewidth]{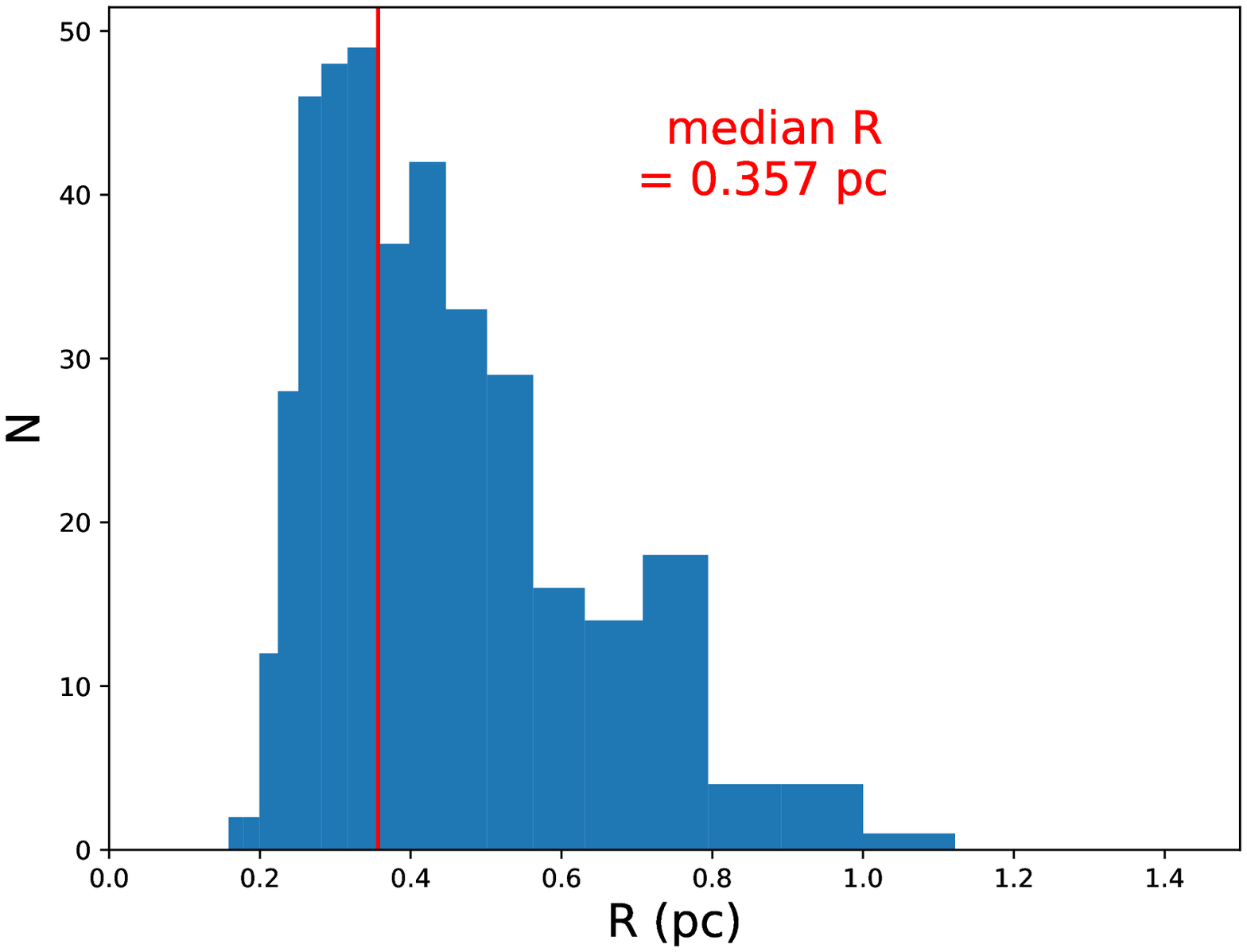}
		\caption{Radius $R$}
		\label{fig:core_R}
	\end{subfigure}
	
	\begin{subfigure}{0.45\linewidth}
		\centering
		\includegraphics[width=\linewidth]{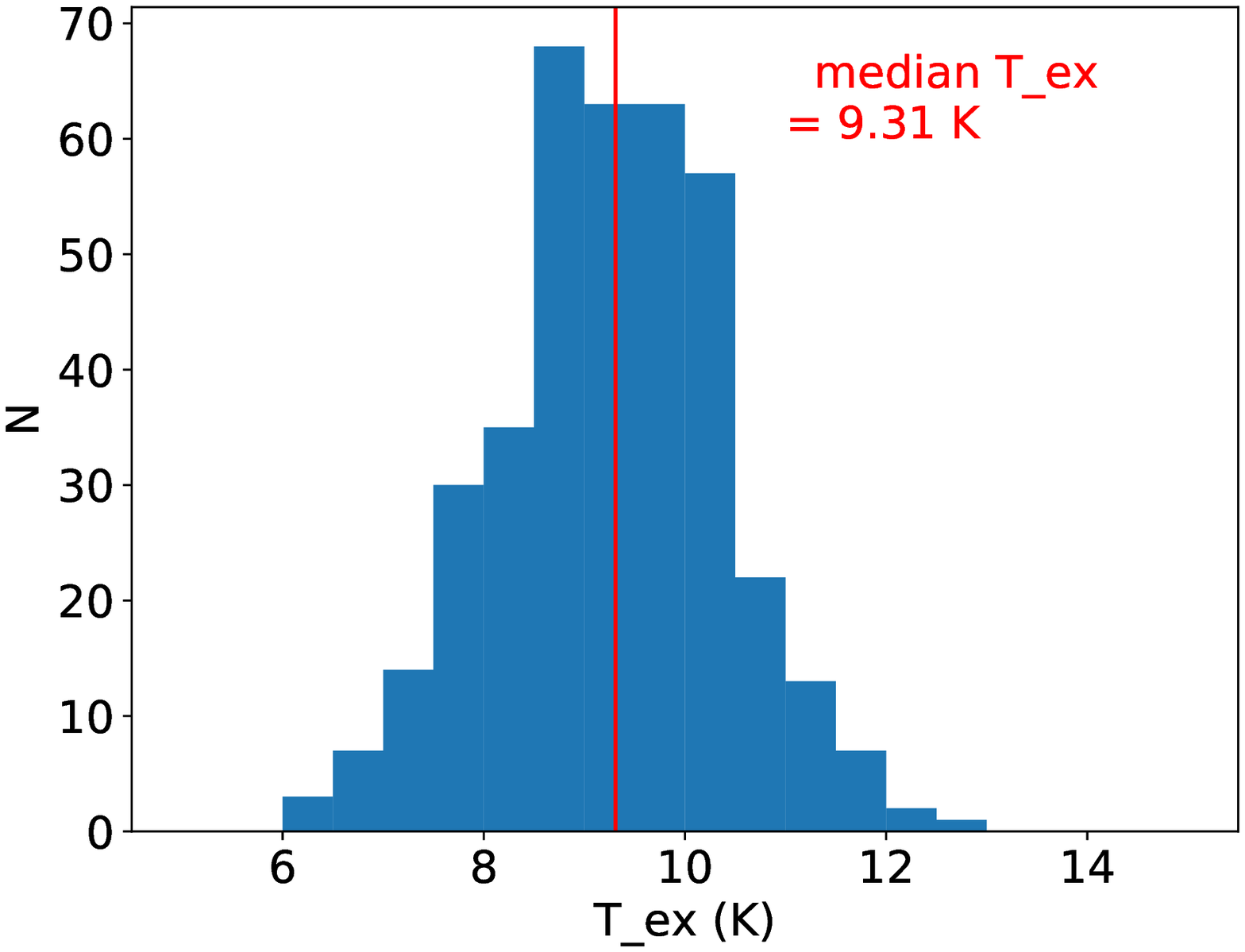}
		\caption{Excitation temperature $T_{ex}$}
		\label{fig:core_Tex}
	\end{subfigure}
	\begin{subfigure}{0.45\linewidth}
		\centering
		\includegraphics[width=\linewidth]{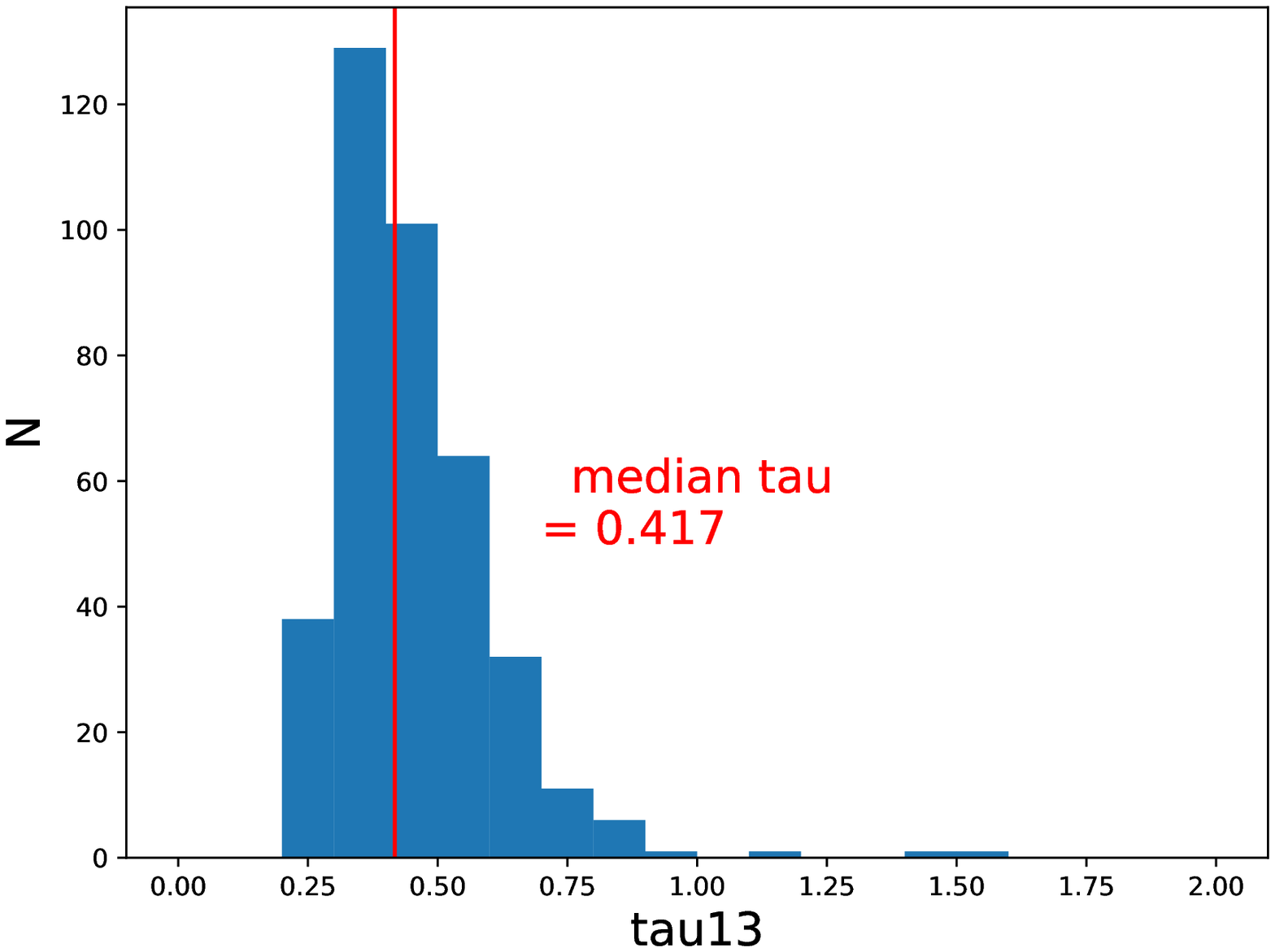}
		\caption{$^{13}$CO optical depth $\tau_{^{13}\mathrm{CO}}$}
		\label{fig:core_tau}
	\end{subfigure}
	
	\begin{subfigure}{0.45\linewidth}
		\centering
		\includegraphics[width=\linewidth]{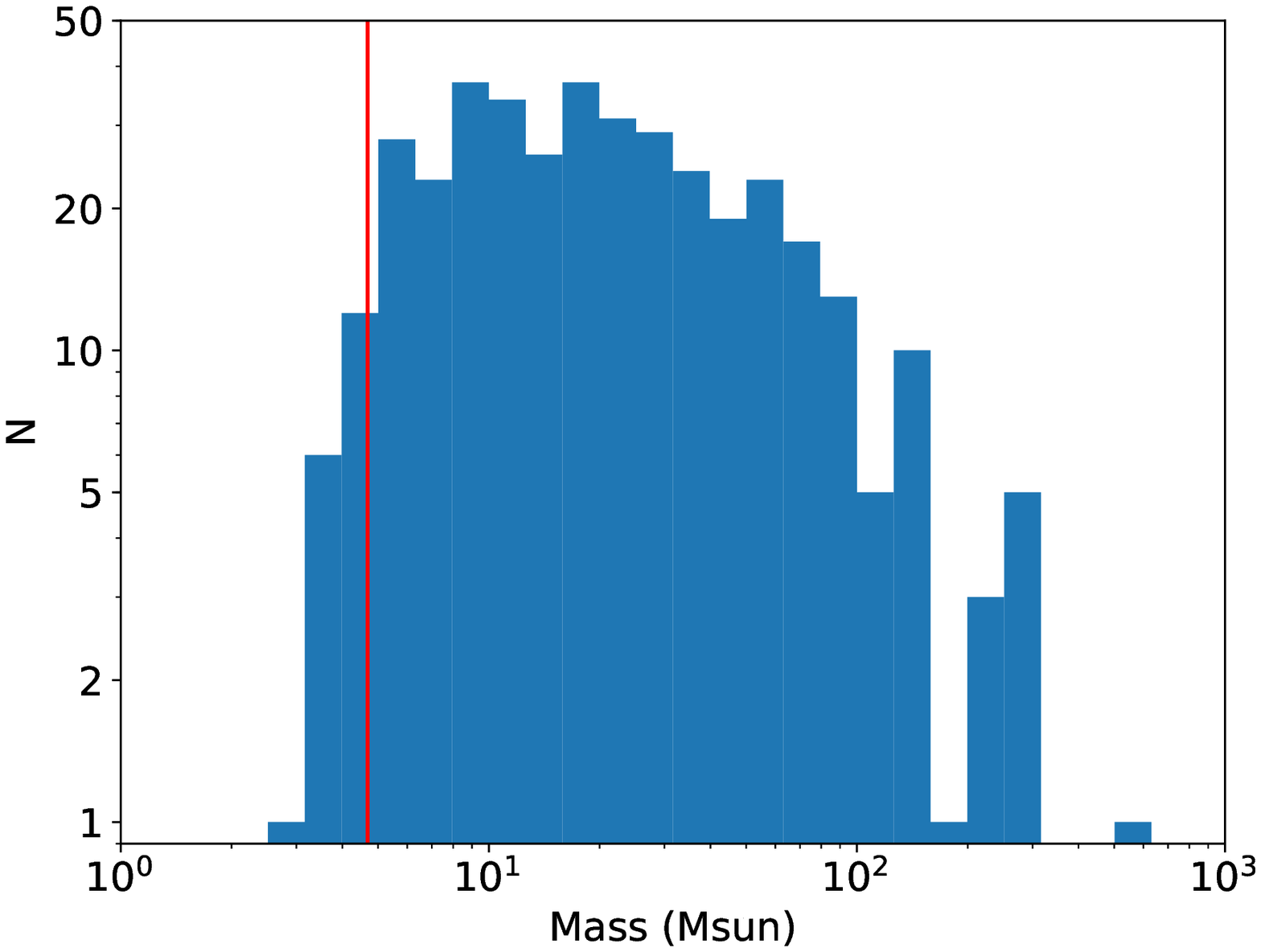}
		\caption{Core mass $M(LTE)$}
		\label{fig:core_CMF}
	\end{subfigure}
	\begin{subfigure}{0.45\linewidth}
		\centering
		\includegraphics[width=\linewidth]{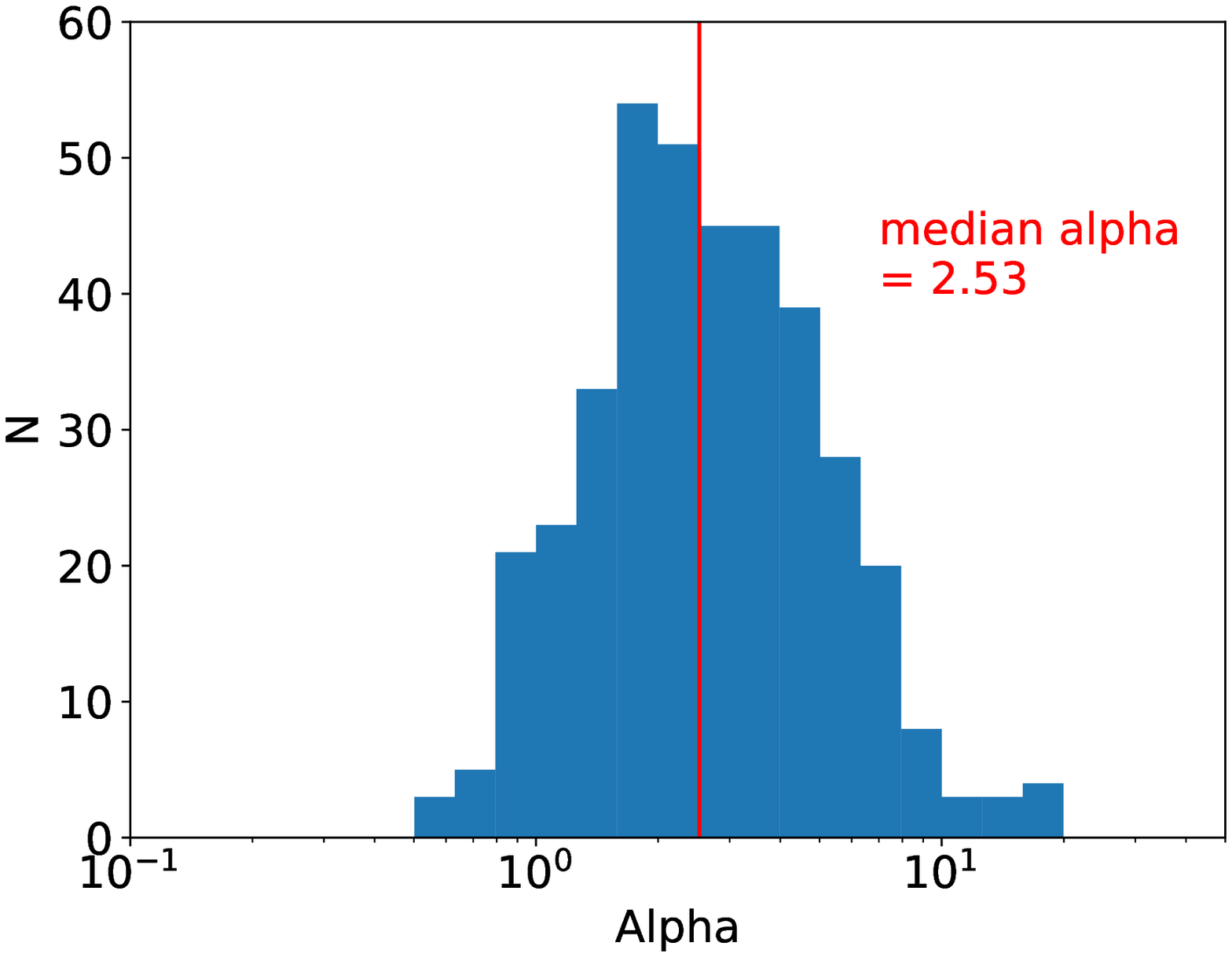}
		\caption{Virial parameters $\alpha$}
		\label{fig:core_Alpha}
	\end{subfigure}
	\caption{Distribution of (a) velocity dispersion of $^{13}$CO emission, (b) $1\sigma$ radius, (c) excitation temperature, (d) $^{13}$CO optical depth, (e) mass, (f) virial parameter for all cores in the entire GMF. Red lines represent median value but a completeness limit in (e). The completeness limit is calculated from a hypothetical core with $4\sigma$ (1.2 K) $^{13}$CO emission in $min\_npix=30$ pixels, while $T_{ex}$ and $\tau_{^{13}\mathrm{CO}}$ are taken to be the median value of all cores.}
	\label{Fig:Para_core}
\end{figure}
%------------------
$^{13}$CO cores in each component are searched using $Python$ package $astroDendro$ \citep{2008ApJ...679.1338R} and we use the minimum structures as the cores.
The settings are $min\_npix = 30, min\_value = min\_delta = 0.6$ K.
It means that the cores we find will include 30 voxels at least, the intensity of each voxel should be greater than 0.6 K (two times of noise), and two contiguous voxels with intensity difference below 0.6 K will not be considered as belonging to different structures.
These values are selected to let the resultant cores be real and robust against noise.
% fitted parameters
The $1\sigma$ radius $R$ and velocity dispersion $\sigma_{v, ^{13}CO}$ are calculated automatically from the moment method and without deconvolution by the beam size.
Excitation temperature for each core is derived using the peak $^{12}$CO intensity within the corresponding voxels with equation \ref{eq:T_ex_value}.
The peak optical depth of $^{13}$CO is derived using peak $^{13}$CO intensity within the corresponding voxels of each core with equation \ref{eq:tau_value}. 
We derived the mass of each core using the LTE method with equation \ref{eq:N_H2(LTE)} and the mean molecular weight per H$_2$ molecule is $\mu_{H_2} = 2.8$ \citep{2008A&A...487..993K}.
The total flux is calculated from the voxels that define in the cores, without any extrapolation.
Distributions for these physical parameters of the cores are shown in Figure \ref{Fig:Para_core}.
It shows that those cores have masses of $10 \sim 100 M_\odot$, sizes of $\sim$0.3 pc and excitation temperatures of about 9 K.

% Different Core Mass among Each Part
% Cloud Mass and Total Core Mass
\begin{table}
	\begin{center}
		\caption{Cloud Mass and Total Core Mass of each Component}
		\begin{tabular}{crrrrrrr}
			\hline\noalign{\smallskip}
			Part & $M(X_{CO})$ & $M(LTE)$ & $M(cores)$ & $N(cores)$ & $\dfrac{M(LTE)}{M(X_{CO})}$ & $\dfrac{M(cores)}{M(X_{CO})}$ & $\dfrac{M(cores)}{M(LTE)}$			\\
			& ($ 10^3 M_\odot$) & ($ 10^3 M_\odot$) & ($ 10^3 M_\odot$) & 		&		&		&	\\
			\hline\noalign{\smallskip}
			F1  &      53  &    22  &       4.6  &         116  &    0.41  &      0.086  &       0.21 \\
			F2  &      67  &    22  &       4.9  &         138  &    0.33  &      0.072  &       0.22 \\
			F3  &      10  &   2.5  &       0.8  &          27  &    0.24  &      0.082  &       0.34 \\
			SWP &      66  &    23  &       4.2  &         137  &    0.35  &      0.063  &       0.18	\\
			Entire  &     202  &    72  &      13.8  &         385  &    0.36  &      0.068  &       0.19	\\
			
			\hline\noalign{\smallskip}
		\end{tabular}
		\label{Tab:Mass}
	\end{center}
\end{table}
The derived masses for cloud and the sum for cores in each component are shown in Table \ref{Tab:Mass}.
F3 has a relative low $\dfrac{M(LTE)}{M(X_{CO})}$ ratio and a high $\dfrac{M(cores)}{M(LTE)}$ ratio, but its $\dfrac{M(cores)}{M(X_{CO})}$ ratio is comparable to others.
This means that F3 is faint in $^{13}$CO line emission and therefore has a relative lower LTE mass, but it has nearly the same mass fraction (about 7\%) of dense cores as traced by $^{13}$CO to the diffuse molecular cloud as traced by $^{12}$CO emission.
Taken together with the morphology, it is possible that F3 is not physically associated with other components of the GMF.

% Virial
The one-dimensional velocity dispersion for free particles is related to that of $^{13}$CO through the following equation,
\begin{equation}
\sigma_v= \sqrt{ \sigma^2_{v, ^{13}CO} + k_B T_{ex}(\dfrac{1}{\mu_p m_H} - \dfrac{1}{m_{^{13}CO}}) }
\label{eq:sigma_v}
\end{equation}
where $\mu_p=2.37$ is the mean molecular weight per free particle \citep{2008A&A...487..993K}, $m_H$ is the mass of H atom and $m_{^{13}CO} = 29~m_H$ is the mass of $^{13}$CO molecule.
The virial mass is derived as $ M_{vir} = 5 \sigma^2_v R_{eff}/G $ where $G$ is the gravitational constant and $R_{eff}$ is the core radius derived from the moment method and deconvolved by the beam size of the telescope.
The virial parameter is $ \alpha \equiv M_{vir} / M(LTE) $ \citep{2013ApJ...779..185K}.
If the virial parameter of a structure is significantly larger than 2, it will expand and dissolve in the future, or need external pressure to be stable.
Alternatively, if the virial parameter is much lower than 2, the structure will collapse due to gravity \citep{2008A&A...487..993K}.
As shown in Figure \ref{fig:core_Alpha}, the median value of virial parameter is about 2.5, showing that the majority of the cores are nearly gravitationally stable.

%-------	alpha~mass
\begin{figure}[]
	\centering
	\includegraphics[width=0.8\linewidth]{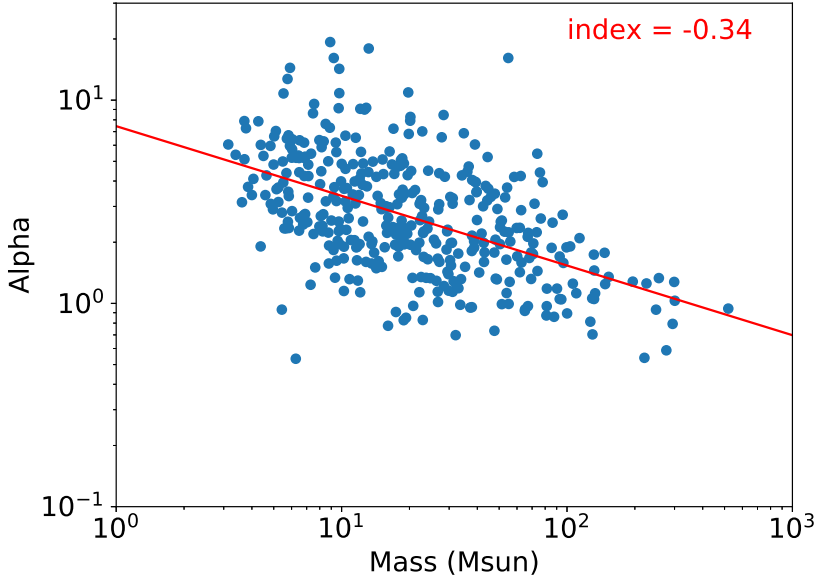}
	\caption{Relation between virial parameter $\alpha$ and LTE mass for each core. The index from power-law fitting is -0.34.}
	\label{Fig:alpha_mass}
\end{figure}
% Difference Virial Parameter among Cores
The relation between virial parameter and core mass is shown in Figure \ref{Fig:alpha_mass}.
It shows that virial parameters of the cores have a power-law relation against their masses with an index of -0.34.
Such a power-law relation is widely observed, for example Figure 1 of \citet{2013ApJ...779..185K}.
The relation is a combination of the mass-size and linewidth-size relations.
The slope is comparable to the theoretical value of -2/3 for pressure-confined fragments in \citet{1992ApJ...395..140B}.

\section{Discussion}
\subsection{Relation between the GMF and spiral arms}
The mean velocity of the GMF is at the center of the velocity range of Sagittarius--Carina arm in longitude-velocity space in the spiral arm model of \citet{2019ApJ...885..131R}.
The dust distance of the GMF from \citet{2018MNRAS.478..651G} is also within the range of Sagittarius–-Carina arm in the model of \citet{2019ApJ...885..131R}.
Hence we place the GMF at the Sagittarius--Carina arm.
However the GMF has a large angle with the Galactic plane, hence we can hardly to call it as the spine of the Milky Way (more like the rib).
We note that, according to Figure 5 of \citet{2019ApJ...885..131R}, the Sagittarius–-Carina arm is relatively thick and extended below the Galactic plane.
The GMF is about 40 pc below the Galactic plane, which is compatible with their results.

\subsection{Candidate of Giant Filament-Filament Collision}
\label{sect:discussion}
We suggest a possible giant filament-filament collision scenario to explain the velocity features of F1 and F2.
Three pieces of evidence are presented below for such a collision.

%-------	L_vpbelt_40
\begin{figure}
	\centering
	\includegraphics[width=\linewidth]{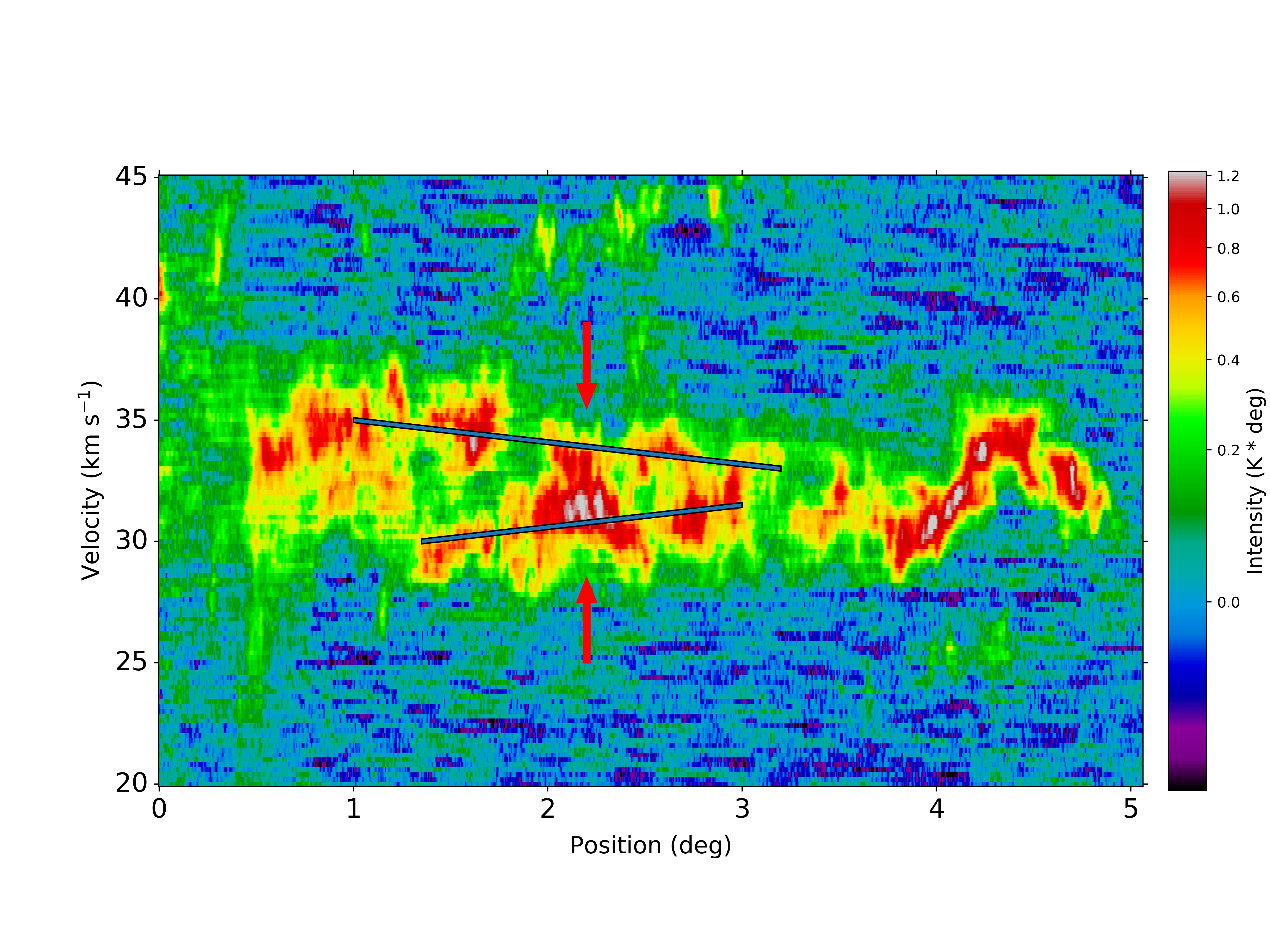}
	\caption{Same as Figure \ref{Fig:L_vpbelt} except for the integral width is 40 pixel ($1/3\dg$). The two lines show the main component of F1 and F2, and the arrows indicate the region of connection between F1 and F2.}
	\label{Fig:L_vpbelt_40}
\end{figure}
%-----------------
A position-velocity map with a narrow integral width of 40 pixels (1$\dg$/3) is shown as Figure \ref{Fig:L_vpbelt_40}.
With a smaller integral range, the connection between F1 and F2 around position offsets 2.2$\dg$ is more clearly shown than in Figure \ref{Fig:L_vpbelt}, as shown by the arrows.
The bridge-like feature is widely used as evidences for collisions between molecular clouds.
If they are overlapped, not physically interacting, then we would see just two main velocity components without the broad bridge feature between them.
The simulations for collision of spherical clouds \citep{2015MNRAS.450...10H} and for collision of filaments \citep{2011A&A...528A..50D} both have verified that the broad bridge feature in the position-velocity diagram is a robust signature of collision.

%-------	HVG
\begin{figure}
	\centering
	\includegraphics[width=\linewidth]{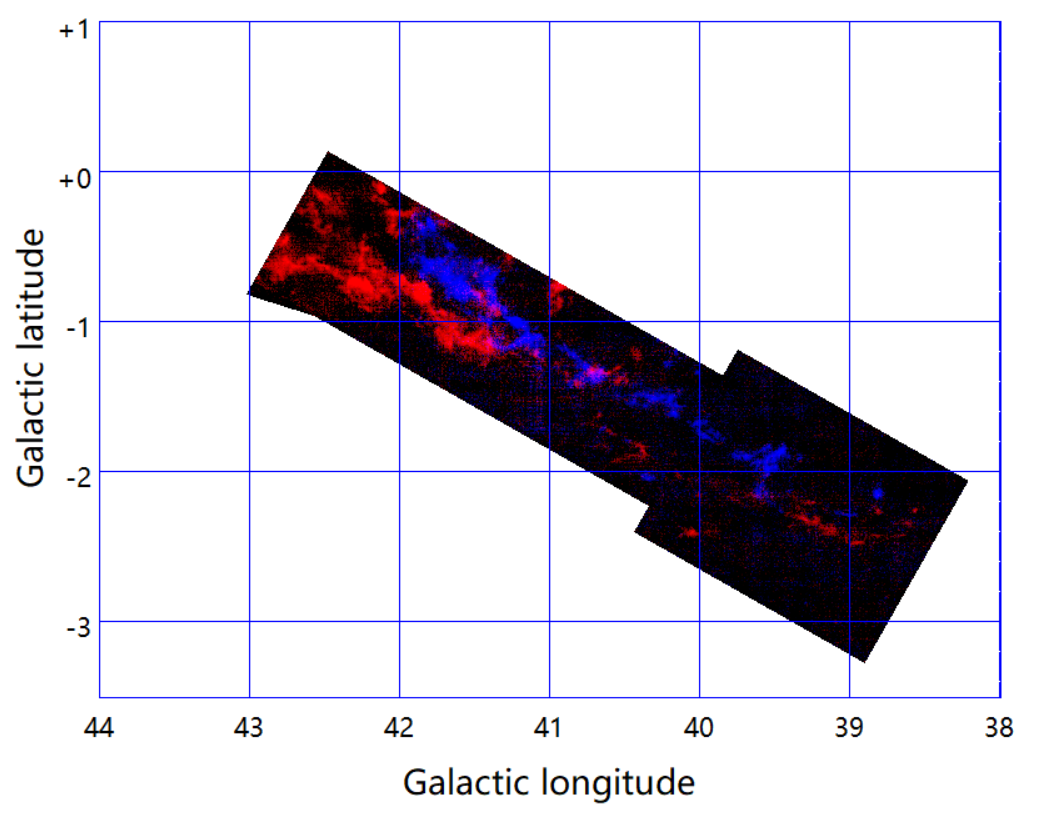}
	\caption{$^{13}$CO high-velocity gases of F1 and F2. Integrated velocity range is 27$\sim$29 km s$^{-1}$ for F1 (in blue) and 35.5$\sim$37.5 km s$^{-1}$ for F2 (in red). Overlaps of HVGs between F1 and F2 in the region around both ($41.5\dg, -0.8\dg$) and ($40.6\dg, -1.4\dg$) are significant.}
	\label{Fig:HVG}
\end{figure}
%-----------------
\citet{2017ApJ...850...23B} have made synthetic observations based on the simulation results of collision between two spherical clouds.
Their results show that overlaps between integrated intensity maps of high-velocity gases (HVGs), i.e. gases with relative velocity higher than one-dimensional velocity dispersion ($\sigma=5$ km~s$^{-1}$ in their simulation), are a clear signature of collision of gas along the line of sight.
We apply the results of this model to the collision between filaments,assuming that collision between different structures (filaments or clouds) will in general have similar physical parameters and results.
The velocity dispersion in F1 and F2 are about 1 km s$^{-1}$, which are much smaller than the value of \citet{2017ApJ...850...23B}.
For our GMF, the velocity ranges for the HVGs are chosen to be between $1\sigma$ and $3\sigma$ from the mean velocity of each component.
Map of $^{13}$CO HVGs in F1 and F2 is shown as Figure \ref{Fig:HVG}, with the integrated velocity range of 27$\sim$29 and 35.5$\sim$37.5 km s$^{-1}$ being shown in blue and red, respectively.
The two HVG parts do overlap in the region around ($41.5\dg, -0.8\dg$) and ($40.6\dg, -1.4\dg$), a total projected length about 20 pc.
The possible collision position is near the centre of both F1 and F2, which is expected as the gravity force in two parallel fluid bars is higher in their centres than the endpoints.

%-------	Tex_F1F2
\begin{figure}
	\centering
	\includegraphics[width=\linewidth]{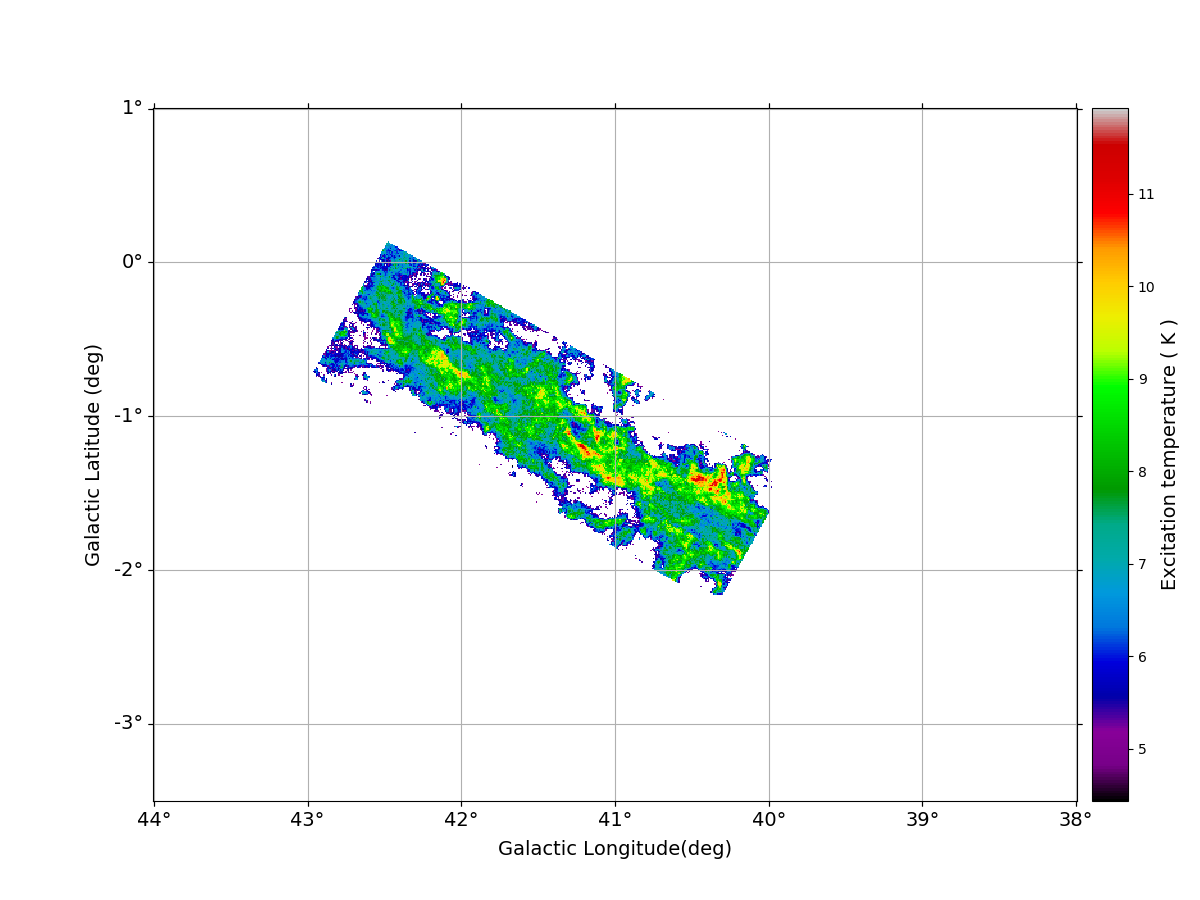}
	\caption{Excitation temperature map of F1 and F2. The collision region roughly from ($41.5\dg, -0.8\dg$) to ($40.6\dg, -1.4\dg$) has higher value than other parts of the GMF.}
	\label{Fig:Tex_F1F2}
\end{figure}
%-----------------
%-------	NH2_F1F2
\begin{figure}
	\centering
	\includegraphics[width=\linewidth]{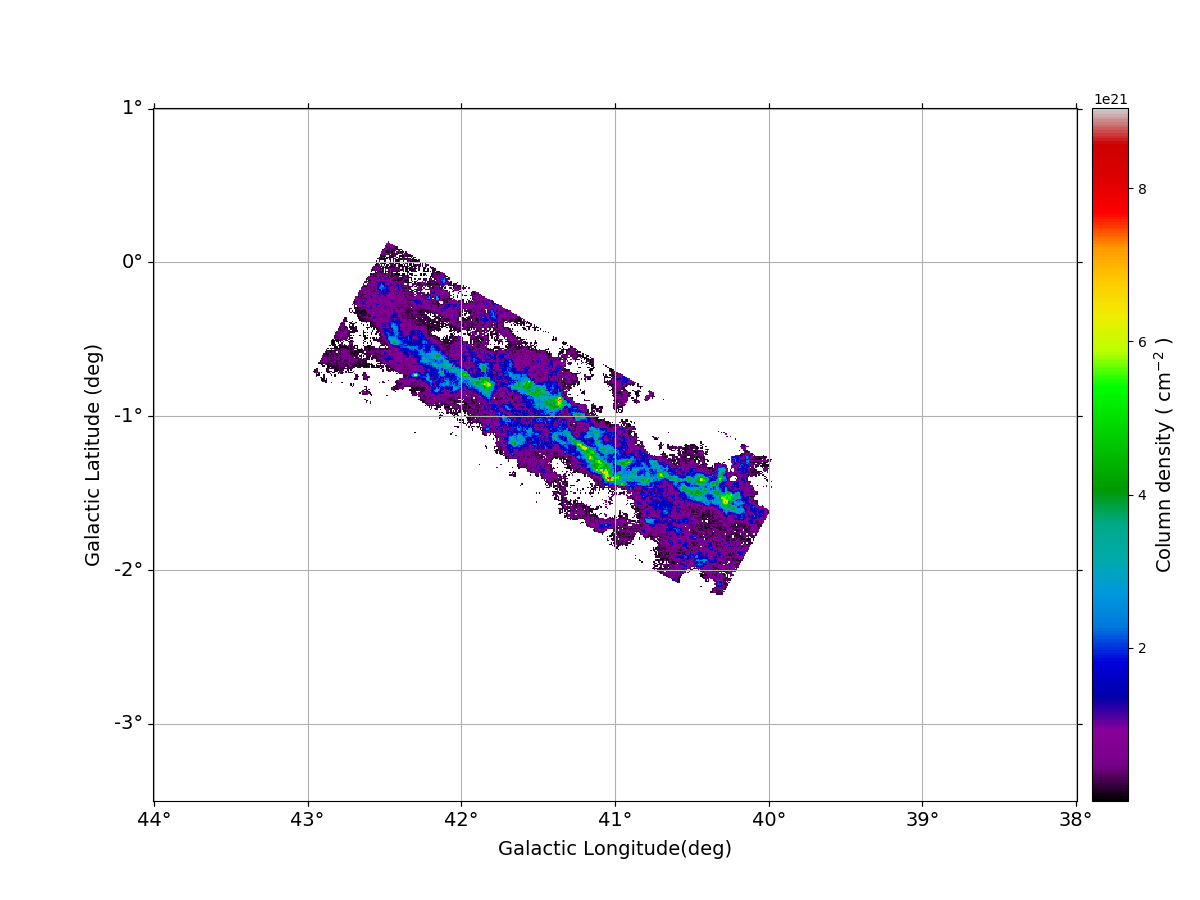}
	\caption{Same as Figure \ref{Fig:Tex_F1F2} except for the column density.}
	\label{Fig:NH2_F1F2}
\end{figure}
%-----------------
A collision usually leads to higher excitation temperatures and column densities in the collision region.
Maps of excitation temperature and column density for F1 and F2 are shown as Figure \ref{Fig:Tex_F1F2} and \ref{Fig:NH2_F1F2}.
It can be seen that the collision region roughly from ($41.5\dg, -0.8\dg$) to ($40.6\dg, -1.4\dg$) is warmer and denser than the nearby non-collision region.

To our knowledge, this is the first candidate for giant filament-filament collision in such a large scale ($\ge10$ pc).
However, \citet{2018MNRAS.479.1722C} have found in their simulations that only about 50 percent of fibers identified in position-position-velocity space correspond to sub-filaments in position-position-position space, whereas the other 50 percent fibers are attributable either to the overlap of several physically  separate sub-filaments or to the contamination from parcels of gas in the line-of-sight.
Therefore, we note that the velocity bridge feature that we have identified in our position-velocity digram of Fig. 11 may be caused by projection effect of sub-filaments or gas parcels in the line-of-sight, rather than a collision of filament-filament.

\section{Summary}
\label{sect:Summary}
Using data from the MWISP project, we have analysed a GMF named GMF MWISP G041-01.
Main results are summarised as follows:

% GMF
(1) The entire GMF has a LSR velocity between 27 and 40 km s$^{-1}$.
At a distance of 1.7 kpc, it has a projected scale of about 160 pc in length and 20 pc in width with a 30$\dg$ angle with respect to the Galactic plane, and has a mass of $2\times10^5 M_\odot$ from the X-factor method.
Its scale and separation from the Galactic plane are unusual when compare to other known GMFs.
Three filamentary components with different velocity are found in the northeast part of the GMF, while the southwest part shows a velocity gradient.

% Para GMF
(2) Physical parameters of the GMF are derived using the LTE method.
The median values are $T_{ex} \approx 7.5$ K, $\tau_{^{13}CO} \approx 0.26$ and $N_{H_2}(LTE) \approx 10^{21}$ cm$^{-2}$, which shows that this GMF is very cold and very diffuse compared to the known GMFs.

% Cores
(3) Cores inside the GMF are searched using $astroDendro$.
Results show that the masses of those cores are $10 \sim 100 M_\odot$ with size of $\sim$0.3 pc and excitation temperature of $\sim$9 K.
Virial parameters of those cores have a median value around 2.5 and are found to have a power-law relation with the masses of cores with an index of -0.34.
For all parts of the GMF, the mass fraction of the total masses of cores to the diffuse cloud traced by $^{12}$CO is around 7\%.

% FFC
(4) A possible giant filament-filament collision scenario is suggested between F1 and F2.
HVGs of these two filaments are overlapped with a range about 20 pc.
The candidate collision position is near the centres of both filaments, showing that a "$\asymp$" shape collision is ongoing.

\begin{acknowledgements}
% normal
The authors thank staffs at Delingha station for their help during the observation, and the anonymous referee for constructive suggestions that improve the manuscript.
L.-H. Lin thanks Fu-Jun Du, Miao-Miao Zhang, Yue-Hui Ma(PMO), Yue-Fang Wu (Peking Univ.) and Tian-Wei Zhang (Univ. of Cologne) for their useful discussion.
% Fund
MWISP project is supported by National Key R\&D Program of China under grant 2017YFA0402700 and Key Research Program of Frontier Sciences of CAS under grant QYZDJ-SSW-SLH047.
H.-C. Wang is supported by the National Natural Science Foundation of China(NSFC) under grant 11973091.
C. Li is supported by the NSFC under grant 11503087.

\textit{Software}: GILDAS-CLASS\footnote{\url{http://www.iram.fr/IRAMFR/GILDAS}}, Astropy\footnote{\url{http://www.astropy.org}} \citep{2013A&A...558A..33A, 2018AJ....156..123A},  astrodendro\footnote{\url{http://www.dendrograms.org/}} \citep{2008ApJ...679.1338R}, Matplotlib \citep{2007CSE.....9...90H} and SAOImage-DS9 \citep{2003ASPC..295..489J}.
\end{acknowledgements}

\bibliographystyle{raa}
\bibliography{RAA-2019-0294R3_export-bibtex}
\label{lastpage}
\end{document}